\documentclass[twocolumn]{aastex63}
\usepackage{booktabs}
\usepackage{gensymb}

\begin{document}

\title{The Masses of Supernova Remnant Progenitors in NGC 6946}

\shorttitle{NGC 6946 SNR Progenitor Masses}
\shortauthors{Koplitz et al.}

\correspondingauthor{Brad Koplitz} 
\email{bmk12@uw.edu}

\author[0000-0001-5530-2872]{Brad Koplitz}
\affil{Department of Astronomy, Box 351580, 
University of Washington, Seattle, WA 98195, USA}

\author[0000-0002-2630-9490]{Jared Johnson}
\affil{Department of Astronomy, Box 351580, 
University of Washington, Seattle, WA 98195, USA}

\author[0000-0002-7502-0597]{Benjamin F. Williams}
\affil{Department of Astronomy, Box 351580, 
University of Washington, Seattle, WA 98195, USA}

\author[0000-0002-4134-864X]{Knox S. Long}
\affil{Space Telescope Science Institute,
3700 San Martin Drive, Baltimore, MD 21218, USA}
\affil{Eureka Scientific, Inc. 
2452 Delmer Street, Suite 100, Oakland, CA 94602-3017, USA}

\author[0000-0003-2379-6518]{William P. Blair}
\affil{Department of Physics and Astronomy, Johns Hopkins University, 
3400 N. Charles Street, Baltimore, MD 21218, USA}

\author{Jeremiah W. Murphy}
\affil{Department of Physics, Florida State University,
77 Chieftan Way, Tallahassee, FL 32306, USA}

\author[0000-0001-8416-4093]{Andrew Dolphin}
\affil{Raytheon Technologies,
1151 E. Hermans Road, Tucson, AZ 85706, USA}
\affil{Steward Observatory, University of Arizona,
933 N. Cherry Avenue, Tucson, AZ 85719}

\author[0000-0002-2895-3028]{Tristan Hillis}
\affil{Department of Astronomy, Box 351580, 
University of Washington, Seattle, WA 98195, USA}

\begin{abstract}

We constrained the progenitor masses for \replaced{175}{169} supernova remnants, 8 historically observed supernovae, and the black hole formation candidate in NGC 6946, finding that they are consistent with originating from a standard initial mass function.
Additionally, there were 16 remnants that showed no sign of nearby star formation consistent with a core-collapse supernova, making them good Type Ia candidates.
Using \textit{Hubble Space Telescope} broadband imaging, we measured stellar photometry of ACS/WFC fields in F435W, F555W, F606W, and F814W filters as well as WFC3/UVIS fields in F438W, F606W, and F814W.
We then fitted this photometry with stellar evolutionary models to determine the ages of the young populations present at the positions of the SNRs and SNe.
We then infer a progenitor mass probability distribution from the fitted age distribution.
For 37 SNRs we tested how different filter combinations affected the inferred masses.
We find that filters sensitive to H$\alpha$, [N II], and [S II] gas emission can bias mass estimates for remnants that rely on our technique.
Using a KS-test analysis on our most reliable measurements, we find the progenitor mass distribution is well-matched by a power-law index of $-2.6^{+0.5}_{-0.6}$, which is consistent with a standard initial mass function.

\end{abstract}


\section{INTRODUCTION} \label{sect1}

Stars with a mass above $\sim$8 M$_{\odot}$ are expected to end their lives as core-collapse supernovae (CCSNe; \cite{Woosley02}).
Many of the heavy elements found in nature are formed and dispersed during this process, enriching the interstellar medium and affecting the star formation of a galaxy.
The number of cataloged SNe is on the order thousands (e.g. \cite{Guillochon17,Holoien19}), however \replaced{only around 30}{few} of these have constraints on the progenitor.

While direct imaging of the progenitor star is the traditional way to determine the mass of SNe progenitors (e.g. \cite{VD03,Smartt03,Smartt04,Li06,GalYam07}), relatively few have been constrained this way.
The most difficult requirement is having spatially resolved photometry (usually better than $\sim$0.1") of the star pre-explosion.
As a result, there have been only \replaced{30}{34} SNe that have had their progenitor masses determined by pre-SN images, with 38 upper limits constrained as well \citep{VD17}\added{\citep{KF18,VD18,ONeill19,Kilpatrick21}}.

Observations of Type II-P SNe show their progenitors are red supergiants with masses $\geq$7 M$_{\odot}$ \citep{Smartt09,Fraser11,Jennings12}, however the upper limit is not known.
Stellar evolution theory suggests a red supergiant upper mass cutoff between 25 and 30 M$_{\odot}$.
\cite{Smartt15} suggests that the upper mass limit for Type II-P progenitors is 17-19 M$_{\odot}$.
\cite{Smartt09} was the first to claim that the lack of high mass progenitors is statistically significant and dubbed it ``The Red Supergiant Problem''.
\cite{DB20} found an upper limit of $19^{+4}_{-2}$ M$_{\odot}$ and suggests the problem may not be as significant as previously thought.
They attribute some of the discrepancy to the small sample size and the steepness of the luminosity-distribution of red supergiants.
Simulations of CCSNe have shown that not all massive stars explode \citep{SW14,Sukhbold16}.
There could be mass ranges in which stars are less likely to explode and instead collapse straight into a black hole \citep{PT15}.
One way to constrain the limits of these islands is to expand the catalog of measured masses of CCSNe progenitors.

To increase the number of progenitor mass constraints we used an age-dating technique on the stellar populations surrounding a SN event.
The method assumes that only massive stars ($>$7.5 M$_{\odot}$) become CCSNe \citep{Jennings12,Diaz18}, leading to a lifespan of $<$50 Myr \citep{Girardi02}.
Previous works have confirmed this experimentally, showing that most young stars within 50 pc were related to the SN remnant (SNR) progenitor \citep{BG06,Gogarten09,Williams14a}.
\cite{Murphy11} showed that the assumption that most young stars within 50 pc of a SNR are related to the progenitor is correct and that the method gives reliable results out to $\sim$8 Mpc.
We can then use the stellar population surrounding a SN event to constrain the age and mass of the progenitor.
In short, the age of a SNe progenitor can be inferred from the color-magnitude diagram (CMD) of the stars in its vicinity.
Using a superposition of stellar populations, we can model the observed CMD and find the most likely age from theoretical isochrones.
By assuming the SN progenitor comes from the youngest statistically-significant population, we are then able to infer the progenitor mass based on stellar evolution models.

We have developed, tested, and applied this technique to determine the progenitor mass of SNRs in the nearby star forming galaxies M31 \citep{Jennings12}, M33 \added{\citep{Jennings14}}\citep{Diaz18}, and M83 \citep{Williams19} \added{, as well as the Small \citep{Auchettl19} and Large \citep{Badenes09} Magellanic Clouds}.
\added{Additionally, we have used this technique to constrain the mass of historically observed CCSNe \citep{Williams14a,Williams18,Diaz21}.}
Here we do the same for the SNR population in NGC 6946.
NGC 6946 is a nearly face-on ($i = 32.6 \degree$, \cite{Blok08}) spiral galaxy that lies near the galactic plane ($b = 11.6 \degree$, \cite{Evans10}).
\replaced{It is currently forming stars at a higher than normal rate, as low as 3.2 M$_{\odot}$ per year \citep{Jarrett13} and as high as 12.1 M$_{\odot}$ per year \citep{Eldridge19}.}{It is currently forming between 3.2 and 12.1 M$_{\odot}$ per year \citep{Jarrett13,Eldridge19}.}
This high rate has caused ten SNe to be observed in the galaxy since 1917, when the average star forming galaxy is expected to host one per century \citep{AS13}, leading to the galaxy being referred to as ``the Fireworks Galaxy".
In 2015, the galaxy was the location of the first observed black hole formation candidate \citep{Gerke15,Adams17}.
Follow up analysis of the stellar population near the disappearing star pointed to the progenitor having an age of $\sim$10.6 Myr \citep{Murphy18}.
\replaced{It was assumed that the star did not have a companion and thus had an initial mass of $\sim$18 M$_{\odot}$.}{Assuming a single star population, the source is thought to have an initial mass of $\sim$18 M$_{\odot}$.}

In this paper, Section \ref{sect2} details the \textit{Hubble Space Telescope} ($HST$) data used, as well as the analysis technique employed to estimate progenitor masses.
Section \ref{sect3} shows the resulting progenitor mass and age estimates.
Section \ref{sect4} examines the distribution of masses in the context of the progenitor mass function as well as quantifying how extended gas emission affects the inferred masses.
Section \ref{sect5} provides a short summary of our conclusions.
We adopt a distance modulus consistent with \cite{Anand18} of 29.4 (7.59 Mpc), which provided slightly better fits (see Section \ref{sect3}) to our data than their central value of 29.45 (7.76 Mpc).

\section{DATA \& ANALYSIS} \label{sect2}

\begin{figure}
    \centering
    \epsscale{1.2}
    \plotone{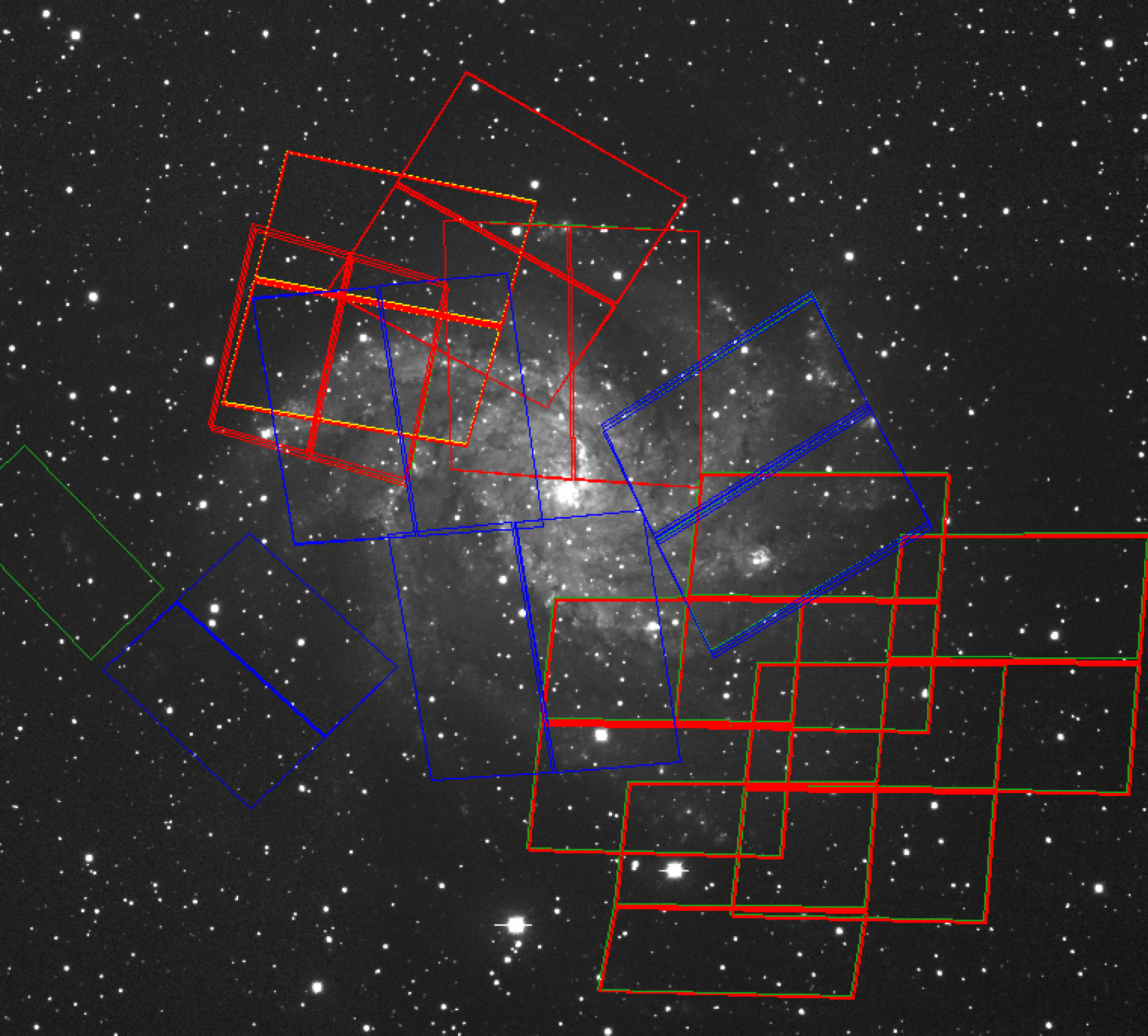}
    \caption{$Hubble$ pointings overplotted on a V band image from Kitt Peak National Observatory's 2.1m telescope. B band pointings are shown in blue, V band in yellow, Broad-V band in green, and I band in red.}
    \label{fig1}
\end{figure}

We analyze $HST$ observations made using the Advanced Camera for Survey (ACS) instrument with Wide-Field Channel (WFC) F435W, F555W, F606W, and F814W filters and the Wide-Field Camera 3 (WFC3) instrument with Ultraviolet-Visible channel (UVIS) F438W, F606W, and F814W filters.
The specific observations that we included in our analysis, along with their exposure details, are listed in Table \ref{tab1}.
The sum of our observations cover most of the main body of the disk of NGC 6946 and can be seen in Figure \ref{fig1}.
We were not able to gather data around each SNR in every filter, therefore some locations were fit using data collected with ACS and some with WFC3.
For simplicity in the text we define ACS/WFC F435W and WFC3/UVIS F438W as B band filters, ACS/WFC F555W as the V band filter, ACS/WFC F606W and WFC3/UVIS F606W as Broad-V band filters, and ACS/WFC F814W and WFC3/UVIS F814W as I band filters.
Table \ref{tab2} indicates which camera-filter combinations were used to fit individual remnants.

Below, we first discuss our catalog of SNR locations.
Then we describe our technique for measuring and fitting the stellar photometry to obtain age estimates, followed by how we constrain progenitor masses from these ages.

\subsection{Supernova Remnant Locations} \label{sect2.1}

The objects we identify as SNRs (or more probably SNRs) were identified by \cite{Long19} or \cite{Long20}, hereafter L19 and L20 respectively.
The L19 candidates were identified on the basis of high [SII:Ha] ratios compared to HII regions.
The L20 candidates were identified by their strong [Fe II] emission compared to Pa$\beta$.
\added{Using both methods to locate likely SNRs improves the completeness in dusty regions.
See L20 for a full comparison between the two identification methods.}
The full catalog is provided in Table \ref{tab3}.
Column 1 is the SNR identifier from their respective catalog.
Locations in both catalogs are given the L19 name in column 1 with column 2 showing the L20 identifier.
Locations in only one catalog will have column 2 be empty.
Columns 3 and 4 are the RA and Dec (J2000) of the SNR.
Column 5 indicates the use flag for the SNR.

The use flag indicates how the SNR was or was not used in our analysis.
We separate our locations into the following categories:
\begin{itemize}
    \item p: Progenitor mass constrained from measured local young population.
    \item h: Progenitor mass constrained by measured local population, but with filters that include some emission lines (see Section \ref{sect2.6}).
    \item n: No young population detected in the fit (Type Ia candidate).
    \item i: Insufficient photometry for a measurement; fitting failed.
    \item o: Outside of the HST coverage; no measurement performed.
\end{itemize}

There are 225 SNRs, 10 historically observed SNe, and the black hole formation candidate in our catalog.
\replaced{Due to a lack of deep homogeneous coverage in multiple bands, we were able to constrain the star formation history (SFH) for 200 of these.}{Of these, 195 had sufficiently deep imaging in multiple bands for us to constrain the star formation history (SFH).}
\replaced{We did not have any coverage for 35 remnants and one (L20-076) in which we only had I band images and thus could not preform a fit.}{We were unable to perform a fit for 35 SNRs because they reside outside of our $HST$ coverage. For the remnant L20-076, we were only able to obtain I band images and thus could not make a CMD to perform a fit.}
\replaced{For five locations, including SN2004et, we didn't recover enough of a stellar population  for a fit to converge.}{For the historical SN SN2004et and four SNRs, we recovered only 1 or 2 stars within 50 pc which was insufficient to provide a population age constraint.}
We were able to gather data in \replaced{more than two}{three} filters for 37 SNRs and \replaced{6}{5} observed SNe.

A control group of 173 regions was also generated throughout the galaxy.
\added{The positions of these were chosen to be 100 pc away from an SNR, in a random direction, to mimic the spatial distribution of the SNR catalog while not fitting the same stellar populations.}
This set of random locations were \deleted{chosen to mimic the spatial distribution of the SNR catalog and was} used to verify that the SNR results were different than that of a random sample.
The group was run through the same measurement process as the SNR catalog (see Section \ref{sect4.2} for details).

\subsection{Photometry} \label{sect2.2}

We measured resolved stellar photometry using the point spread function (PSF) fitting pipeline that was used for the Panchromatic Hubble Andromeda Treasure (PHAT; \cite{Williams14b}).

In short, we put all of the exposures listed in Table \ref{tab1}, after bad pixel flagging with {\tt astrodrizzle} through the DOLPHOT \citep{Dolphin00,Dolphin16} photometry routines.
Using the DOLPHOT parameters updated from those of the PHAT survey to optimize the resulting catalogs for stellar populations science.
The main updates are the removal of catalog-level CTE corrections, because we used the on-image CTE corrections ({\tt{flc}} images), and the use of TinyTim PSFs \citep{Krist11} for all cameras and filters.

Memory and time limitations prevent us from simply putting the entire set of exposures into DOLPHOT simultaneously.
Instead, we subdivided the data into separate stacks to measure the photometry of different regions of the survey in parallel.
We used DOLPHOT parameters that allow the user to define the region within which it performs photometry to launch multiple photometry processes, each with a different region of the survey including all overlapping individual images.
We made these regions sufficiently small that DOLPHOT could complete the PSF fitting photometry in a few days.
We set up 16 separate processes, each covering a fraction of the survey area, overlapping by 100 pixels on a side to avoid introducing edge effects.
We then merged the resulting catalogs along the centers of the regions' overlaps to produce one final catalog for the survey.
This full catalog was then processed to flag poor measurements, as described below.

\subsection{Flagging and Processing Photometry Output}\label{sect2.3}

DOLPHOT returns a comprehensive table of all of the measurements made on every PSF fit to every image, as well as the combined measurement of every source in every filter.
These measurements include the flux, Vega system magnitude, count-based uncertainty, \texttt{signal-to-noise} ratio, and several measurements of how well the source was fitted by the PSF.
These quality metrics include \texttt{sharpness}, \texttt{roundness}, $\chi$, and \texttt{crowding}.
Full descriptions of these are included in the DOLPHOT documentation\footnote{\url{http://americano.dolphinsim.com/dolphot/}}.
Briefly, the \texttt{sharpness} parameter measures how centrally peaked the source is compared to the PSF.
High values signify too centrally-peaked, such as a hot pixel or cosmic ray.
Low values indicate that the source is not peaked enough, as expected for blended stars or background galaxies.
The \texttt{roundness} parameter measures how circular the source is, and $\chi$ provides an estimate of the overall goodness of fit to the PSF.
The \texttt{crowding} parameter measures how much the source's photometry is affected by neighboring sources.
The larger the crowding value, the more densely packed the PSF radius is with other sources, and the more likely it is that the reported magnitude has systematic uncertainties due to subtraction of neighbors.

We tested a grid of possible cuts on these parameters to optimize the quality of our final photometry catalog.
We determined that the \texttt{signal-to-noise}, \texttt{sharpness}, and \texttt{crowding} parameters affected the quality of our catalog most significantly.
We compared the CMD of the stars kept to one of the stars removed by each set of cuts.
We looked for the set that had no CMD features in the stars cut while removing the most stars outside of features in the stars kept.
These tests show that the optimal cuts for our photometry are with \texttt{signal-to-noise} ratios greater than 4.5, \texttt{sharpness} squared values less than 0.4, and \texttt{crowding} values less than 2.
These were the values at which the quality of our photometry was consistent even when small variations to the cutting parameters were introduced.
As seen in Figure \ref{fig2}, the resulting CMDs show well-defined features with little contamination, showing the effectiveness of our quality metrics.

\subsection{Running MATCH} \label{sect2.4}

To derive SFHs for each SNR, we used the CMD fitting program MATCH \citep{Dolphin02,Dolphin13}.
The package fits the CMDs of resolved stars near each SNR to infer the most likely SFH in the region.
The output SFH is a distribution of ages and metallicity of the stellar population within the region, including the uncertainties on the measurements.
MATCH has been used numerous times to measure ages of SNe progenitors (e.g. \cite{Jennings12,Jennings14,Williams14a,Williams19}, star cluster ages (e.g. \cite{Senhyna15,Weisz14,Gossage18}), as well as the SFH for nearby galaxies (e.g. \cite{Williams09,Skillman17}).
During our analysis we applied time bins from 6.6 to 8.0 log years in 0.05 dex steps, and then from 8.0 to 10.2 log years in 0.1 dex steps.
By including the older ages we avoid the issue of the fitting code assigning old stars to young populations.

\subsubsection{Artificial Star Tests} \label{sect2.4.1}

MATCH requires artificial star tests (ASTs) to apply bias and uncertainty to the models for proper comparison to the data.
These tests add a star of known brightness and color to the images at a given location, then reruns the photometry in that region.
This is performed 50,000 times within a 50 pc radius of \added{the claimed location of} a SNR \added{, columns 3 and 4 from Table \ref{tab3}}.
ASTs are computationally expensive, as they require determining photometric measurements multiple times.
Thus, we did not perform a full set at each SNR location.
Instead we ran the full sets at representative locations throughout the data set to generate a library of ASTs that we could use for every SNR, as described in more detail below.

To determine an appropriate grid of locations for our AST library, we measured the relative depth and density for each region.
Due to the differing exposure times present across our data, we only considered magnitudes where the completeness was high ($>$90$\%$) for the entire galaxy.
We then determined the number of stars brighter than our magnitude limit (Broad-V$<$26) divided by the area of each region (stellar density) as well as the faintest star with a \texttt{signal-to-noise} ratio greater than 4.5 (photometric depth) within 50 pc of each SNR.

To determine the number of ASTs we needed to perform, we \replaced{ran a set at one location and at another of similar depth and density}{created sets at the location of two SNRs of similar depth and density}.
We then fit the CMD of one location with both sets to determine the impact of the AST sample on the resulting SFH.
\replaced{We determined that one set of ASTs could be used for locations with similar depth (within 0.5 magnitudes) and density (within 0.86 stars per sec$^{2}$), and still return consistent age estimates.}{Both fits resulted in consistent age estimates so long as the ASTs were ran for a location of similar depth (within 0.5 magnitudes) and density (within 0.86 stars per sec$^{2}$).}
Based on these results, we were able to cover our full grid of depth and density space with 43 locations.
Figure \ref{fig3} shows the distributions of depth and density for our catalog, specifying which locations we ran ASTs for.
This set allows us to account for the different filter combinations required to fit all of the SNRs.
\cite{Williams17} used a similar technique to reduce the number of ASTs needed for the large area their observations covered.

\replaced{That we are able to obtain reliable results for the historical events shows the validity of our AST assumptions.}{To test these results, we fit the historical events with ASTs created for SNR locations.}
In \cite{Williams14a}, \cite{Williams18}, and \cite{Murphy18}, a set of ASTs were made for each location.
We used sets of ASTs from \added{SNR} locations of similar photometric depth and stellar density.
This allowed us to constrain the progenitor mass for 9 of the historical events even though each did not have its own set of ASTs.
This shows our technique for ASTs can be used in the future to reduce the number of necessary sets.

\subsubsection{Foreground \& Differential Extinction} \label{sect2.4.2}

SNRs are often associated with young populations, which often reside in dusty regions.
Furthermore, NGC 6946 is near the galactic plane ($b = 11.6 \degree$), making the foreground extinction significant ($A_{v}=0.94$; \cite{SF11}).
Since main sequence stars have well defined colors we can assume that the observed extinction comes from the dust.
To account for this, MATCH applies extinction to CMDs as it attempts to fit the population.
There are two types of extinction that are considered: the overall affect across the entire region, $A_{v}$, and the spread of extinction from the distribution of stars along the line of sight, $dA_{v}$.
$A_{v}$ is first applied to each star in the CMD models, which changes where features such as the red-giant branch appear.
Spreading along the reddening line is simulated using $dA_{v}$, which affects the sharpness of features and is determined from the observed widening of the main sequence.
\added{By default, MATCH assumes a maximum differential reddening for the youngest stars ($<$40 Myr old) of 0.5 which linearly decreases for older populations.
$dA_{v}$ is then added on top of this default differential reddeing.}

We determine the $A_{v}$ and $dA_{v}$ values by fitting for a range of possibilities at each location.
We allowed $A_{v}$ to range from 0.5 to 2.4 in steps of 0.05.
We allowed $dA_{v}$ to range from 0.0 to 2.6 in steps of 0.2.
To see how $dA_{v}$ changes our inferred mass, see Figures \ref{fig4} and \ref{fig5}.
MATCH outputs the best maximum-likelihood values of $A_{v}$ and $dA_{v}$ for each location.
Smaller $dA_{v}$ step sizes were tested but the goodness of fit values in the upper middle panel of Figures \ref{fig4} and \ref{fig5} did not vary significantly around the best value when steps were smaller than 0.2.

Our regions, on average, returned an $A_{v}$ value of 1.35 which is higher than the value from \cite{SF11}.
The $A_{v}$ value MATCH returns accounts for reddening in both the Milky Way and NGC 6946, while \cite{SF11} only accounts for the Milky Way, thus it is understandable that our $A_{v}$ values are, on average, higher.
\added{Most of our regions returned a $dA_{v}$ value of 0, suggesting the assumed maximum amount differential reddening added to the youngest populations was sufficient to provide good fits to the data.}

\subsubsection{Chemical Evolution} \label{sect2.4.3}

Using the \textit{zinc} flag within MATCH allows us to constrain the fitted metallicity to increase with time, which can be required when the photometry is not sufficiently deep to directly measure the age-metallicity relationship.
Giving ranges for the initial and final metallicity limits the models that MATCH tries to fit.
We constrained the modern epoch metallicity to $-0.5 \leq$ [Fe/H] $\leq 0.1$.
This range is consistent with previous measurements suggesting a solar metal content for the galaxy \citep{Ced12,Gusev17,Long19}.
\replaced{Our results point to the current gas phase metallicity of NGC 6946 being roughly solar, which is consistent with its mass and morphology.}{The distribution of the metallicities from our results were roughly even across the allowed range. 
This is expected given that young populations are not sensitive to metallicity.}

\subsubsection{Contamination CMD} \label{sect2.4.4}

MATCH allows one to include a ``contamination" CMD as part of the fit to isolate populations that are unique to the region being measured.
We generated a contamination CMD by including the stars in an annulus extending from 1.33'' to 1.1' (50 to 1000 pc) which are less likely to be associated with the SNR progenitor.
We then scaled this contamination CMD to the area of our extracted region.
\replaced{Doing so allows MATCH to look for populations present in both samples and compare their densities.
If a population of a specific age has a higher density in the extracted region as compared to the annulus, that would indicate a more likely age for the progenitor.}{By including this background, MATCH will weight any stellar populations unique to the SNR position more heavily than those that are present in the surrounding area of the galaxy.}
This feature was of great use to us for identifying statistically-significant young populations at the locations of the SNRs that were less clustered in the surrounding area.

To determine how a contamination CMD affects the most-likely age, we ran all of our fits both with and without such a contamination CMD.
We found that not including a contamination CMD resulted in systematically lower progenitor mass estimates, likely because of contaminating populations diluting the age distribution (see Section \ref{sect2.5}).
To have the most reliable results we chose to include a contamination CMD when fitting.

\subsubsection{Uncertainties} \label{sect2.4.5}

Fits to CMDs with stellar evolution models have both random and systematic uncertainties.
The random uncertainties mostly come from photometric errors, incompleteness, and the amount of stars used to determine the most likely progenitor age.
The systematic uncertainties are due to our assumption that the models we are fitting are correct.
While in cases with very deep and well-populated CMDs, systematic uncertainties can dominate, in our case where each location has relatively few stars and limited depth, the sampling of the CMD is expected to dominate the uncertainties.
Thus we focus on obtaining reliable estimates on our random uncertainties.
The inhomogeneity of our data causes these to vary between each SNR in our sample, leading us to measure these uncertainties independently for every location.

To determine the random uncertainties, we used the \texttt{hybridMC} task within MATCH \citep{Dolphin13}.
It uses a hybrid Monte-Carlo algorithm that looks at potential SFHs around the best fit and accepts or rejects them based on likelihood.
From the distribution of accepted SFHs, it is able to determine the \added{narrowest} 68\% \replaced{confidence interval for the best fit.}{of the probability that includes the best fit and decreases with lookback time.}
This method for determining the uncertainties is only valid for sets of CMDs that do not share a filter.
For example, B$-$Broad-V and Broad-V$-$I CMDs share the Broad-V filter.
Sharing a filter means the CMDs will be correlated and \texttt{hybridMC} will give incorrect uncertainties.
Thus, for locations with 3 observed bands, we must decide which filter set to use for our mass inference.

We have 236 locations in our catalog.
103 of these could only be fit using a B$-$Broad-V CMD.
53 needed to use a Broad-V$-$I CMD to be fit.
There were two remnants that had to be fit with a V$-$I CMD.
There are 42 locations in our catalog that could be fit with B$-$Broad-V, Broad-V$-$I, or B$-$I CMDs.
To determine which CMD to use for the final fit in these cases, we fit B$-$Broad-V, Broad-V$-$I, and B$-$I CMDs independently.
Using a B$-$I CMD allows us to get the best temperature measurement due to the broader baseline, however fitting each combination allows us to look for systematics in the model fits.
Figure \ref{fig6} shows that mass inferred from a Broad-V$-$I CMD are more evenly spread as a function of mass as compared to the B$-$Broad-V suggesting the Broad-V$-$I distribution is biased towards younger, more massive populations.
B$-$Broad-V CMDs resulted in more masses being inferred between 7.5 and 12 M$_{\odot}$ than the other two distributions suggesting a preference towards older, less massive populations.
Additionally, the B$-$Broad-V and Broad-V$-$I distributions resulted in no Type Ia candidates, which is not likely given the amount of supernova activity in the galaxy.
Thus we used B$-$I CMDs for the 42 locations with data in \replaced{more than 2}{three} filters.
See Section \ref{sect2.6} for a full comparison between the different CMDs as well as why it may be difficult to apply Broad-V band data when inferring SNe progenitor masses using this technique.

\subsection{Progenitor Mass Constraints} \label{sect2.5}

The technique we employed to determine the progenitor mass from our CMD-fitting results has been used in several previous works (e.g. \cite{Jennings12,Williams18,Williams19}).
We limited the age range we considered in our analysis to $<$50 Myr, which corresponds to a main sequence turn off of $\sim$7 M$_{\odot}$.
\cite{Jennings12} empirically found this to be the minimum mass for CCSNe progenitors in single star systems.
It has been shown that an older progenitor is possible in binary systems (e.g. \cite{Xiao19}), though \cite{Diaz18} found the same limit even though older ages were analysed.
The stars in the older time bins are less likely to be correlated with the SNR, as they will have had more time to drift away from their parent cluster.
The older populations have a turn off point near our completeness limits, which tends to increase the size of the error bars associated with each measurement.
Thus limiting our age range reduces the likelihood of contamination from other populations.

MATCH returns the best-fit star formation rate and associated uncertainties for each time bin in each region.
As in \cite{Williams19}, we use these rates to calculate the median age of the young ($<$50 Myr) stars.
The uncertainties on this median age are then determined by resampling the measured rates, taking into account the uncertainties, and recalculating the median age thousands of times and determining the 16th and 84th percentiles of the distribution of median ages.
Finally, we map the resulting age and uncertainties to their corresponding masses by assuming the PARSEC stellar evolution models \citep{Bressan12}.

\subsection{Impact of Gas Emission on Ages} \label{sect2.6}

SNRs are typically identified by their bright emission lines.
The brightest among these are the [N II] doublet (6550,6585\AA), [S II] doublet (6718,6733\AA) and H$\alpha$ (6563\AA), all of which fall within the Broad-V bandpass.
Additionally, they can sometimes reside in H II regions where photons from young massive stars have ionized the adjacent gas.
At the distance of NGC 6946, this emission could be measured by DOLPHOT as blended stars.
These regions of excited gas, and the stars residing in them, appear anomalously red in B$-$Broad-V and anomalously blue in Broad-V$-$I, as seen in Figure \ref{fig7}.
Such colors indicate an overestimate of the Broad-V band brightness compared to the other bands, suggesting that there is additional flux, likely due to gas emission, being included in that band.
While such contamination is not expected to be significant in most cases, its impact will be systematic, as it will always cause the Broad-V measurements to be anomalously bright.
Therefore, it is of particular concern for statistical studies of the SNR distribution.

To test for systematic effects from such gas emission on our SFH results, we fit the 37 SNRs with data in \replaced{more than two}{three} filters using B$-$I, B$-$Broad-V, and Broad-V$-$I colors.
Figure \ref{fig8} compares the distribution of progenitor masses based on the choice of filter combination for these remnants.
The Broad-V$-$ CMDs resulted in a more even distribution than the others, while the B$-$Broad-V distribution has no masses between 40 and 50 M$_{\odot}$.
These results show that using a B$-$Broad-V CMD biases our results towards older populations, where a Broad-V$-$I CMD biases towards younger populations.

We compare how the inferred progenitor masses of individual SNRs changed whether B$-$Broad-V and Broad-V$-$I CMDs were used in Figure \ref{fig6}.
Our results suggest that in some cases Broad-V data is biasing progenitor mass estimates.
In 15 of the 37 cases, the Broad-V$-$I measurement was at least 10 M$_{\odot}$ larger than the B$-$Broad-V for the same region.

To investigate the cause of these discrepancies, we looked at the locations of the discrepant SNRs.
One of the most extreme examples was L19-098, which is a particularly bright example of an SNR nebula, likely because it surrounds an ultraluminous x-ray (e.g. \cite{Dunne00,Blair01}), making it much brighter than the other SNRs in the our catalog.
Because of this, it provides the most obvious example of what strong gas emission can do to our progenitor measurements when the Broad-V filter is used.
Another, more typical example was L19-023.
We show color images of these two examples in Figure \ref{fig9}, with B band data being depicted in blue, Broad-V in green, and I in red.
In both examples, the Broad-V$-$I mass measurements were significantly more massive than those from other filter combinations, as would be expected if the Broad-V data were systematically bright compared to the other bands.

Furthermore, in Figure \ref{fig10} we show combined CMDs of the stars near L20-023 (right) and L19-098 (left) that passed our quality cuts in B, Broad-V, and I filters.
The Broad-V$-$I CMDs are shown in blue, B$-$Broad-V in red, and B$-$I in yellow.
In both examples, the stars are measured to be too blue when using a Broad-V$-$I CMD, with L19-098 mostly having colors between -1.0 and -0.5.
The stars around L19-098 are very red when using B$-$Broad-V, with no star having a color $<$0 and most between 1.5 and 2.5.
When using a B$-$I CMD, the stars around both locations have colors between 0.5 and 1.5 typical of main sequence stars in NGC 6946.

\begin{figure}
     \centering
     \epsscale{0.9}
     \plotone{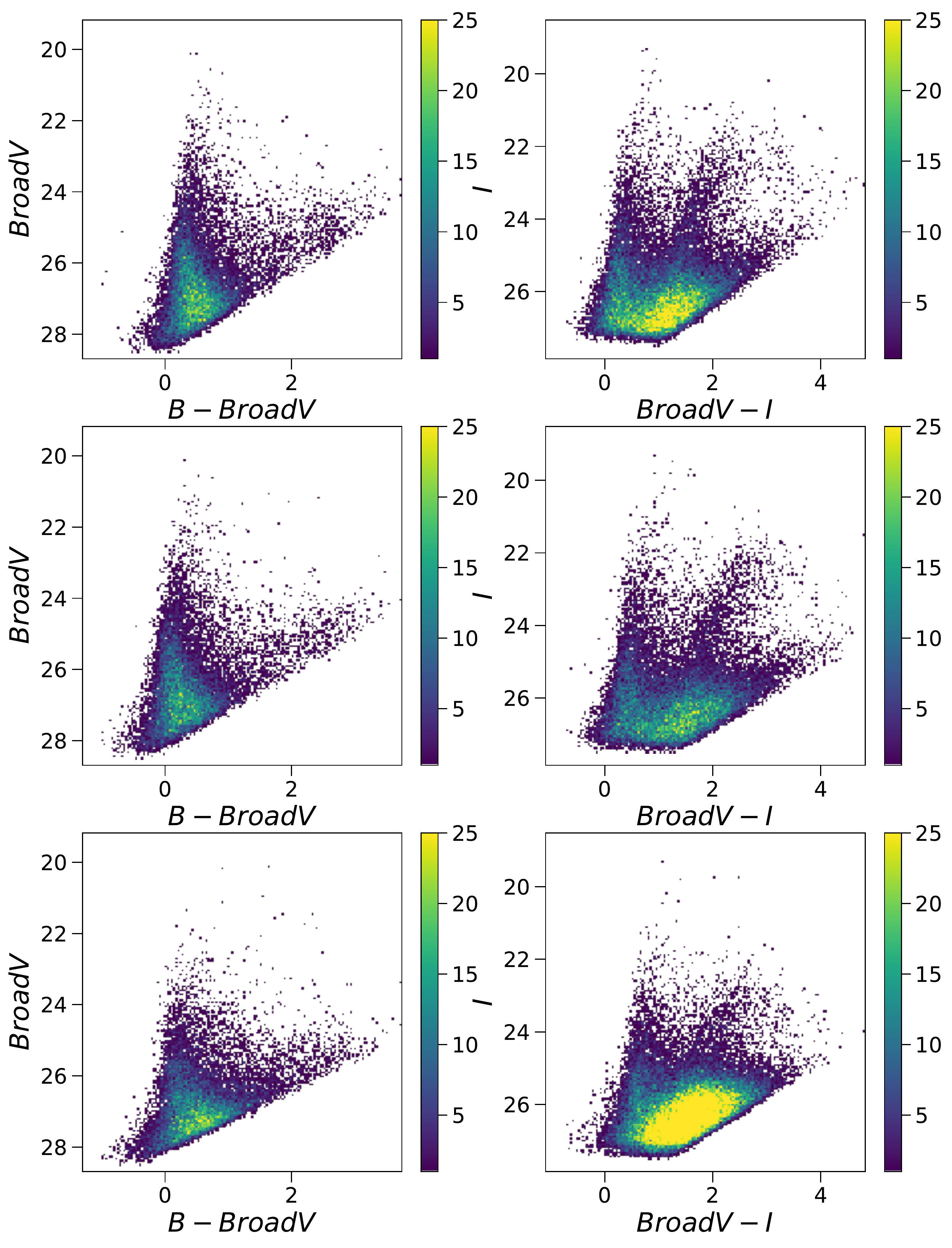}
     \caption{Color-magnitude diagrams of the stellar populations within 1 kpc of three SNRs to show the range in quality of photometry. The left CMDs are the bluer filter combination and the right are redder. The top row surrounds L20-117 which corresponds to higher stellar density ($>$5 arcsec$^{-2}$) and shallow photometric depth (Broad-V$<28.4$). The center row surrounds L20-112 which corresponds to lower stellar density and shallow photometric depth. The bottom row surrounds L20-098 which corresponds to lower stellar density and deep photometric depth.}
     \label{fig2}
 \end{figure}

\begin{figure}
     \centering
     \epsscale{1.1}
     \plotone{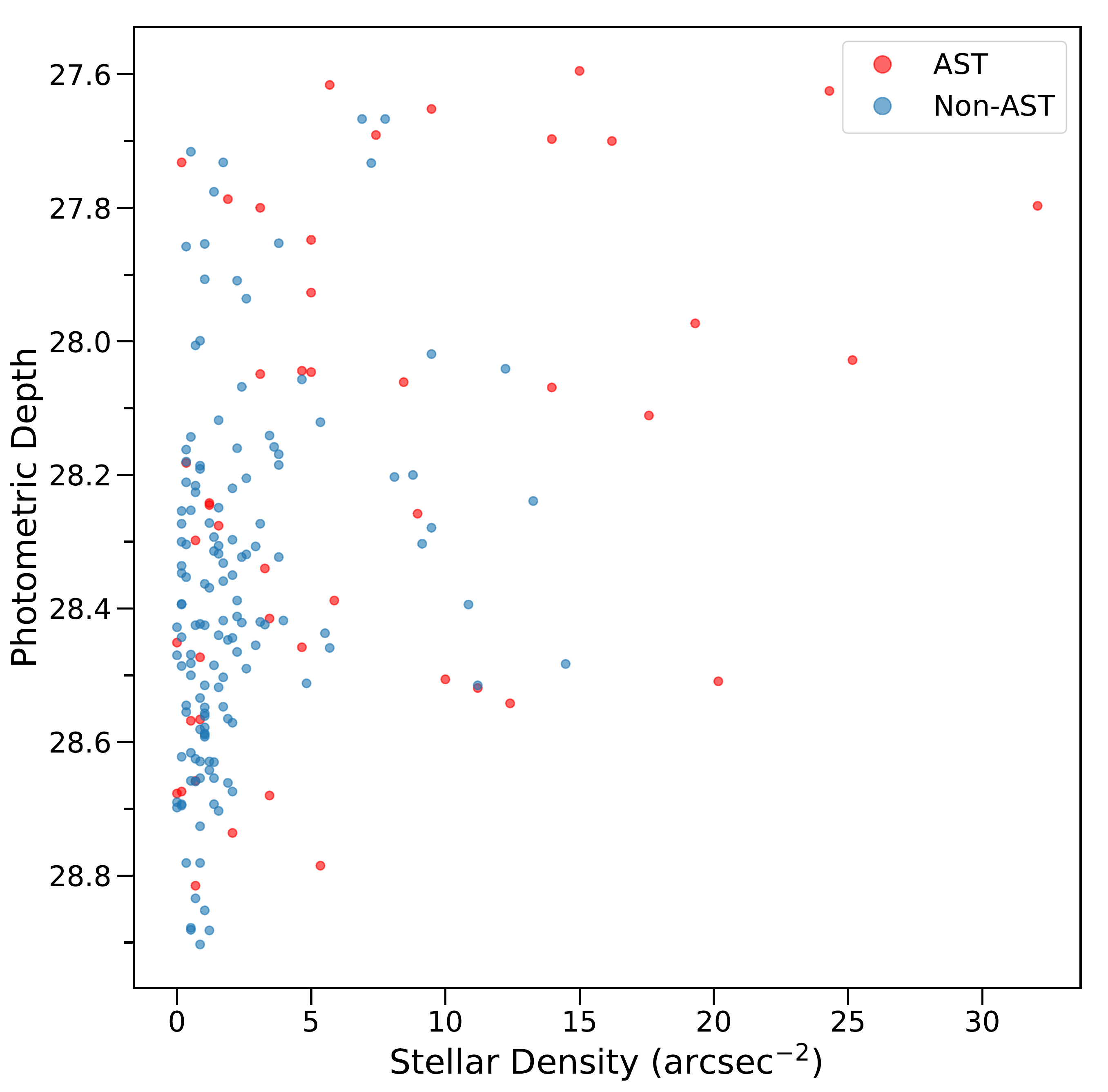}
     \caption{Red dots indicate SNRs where Artificial Star Tests (ASTs) were performed. Blue dots corresponds to SNRs that used ASTs performed at a location within 0.86 in stellar density and 0.5 in photometric depth. See Section \ref{sect2.4.1} for details.}
     \label{fig3}
 \end{figure}

\begin{figure}
     \centering
     \epsscale{1.2}
     \plotone{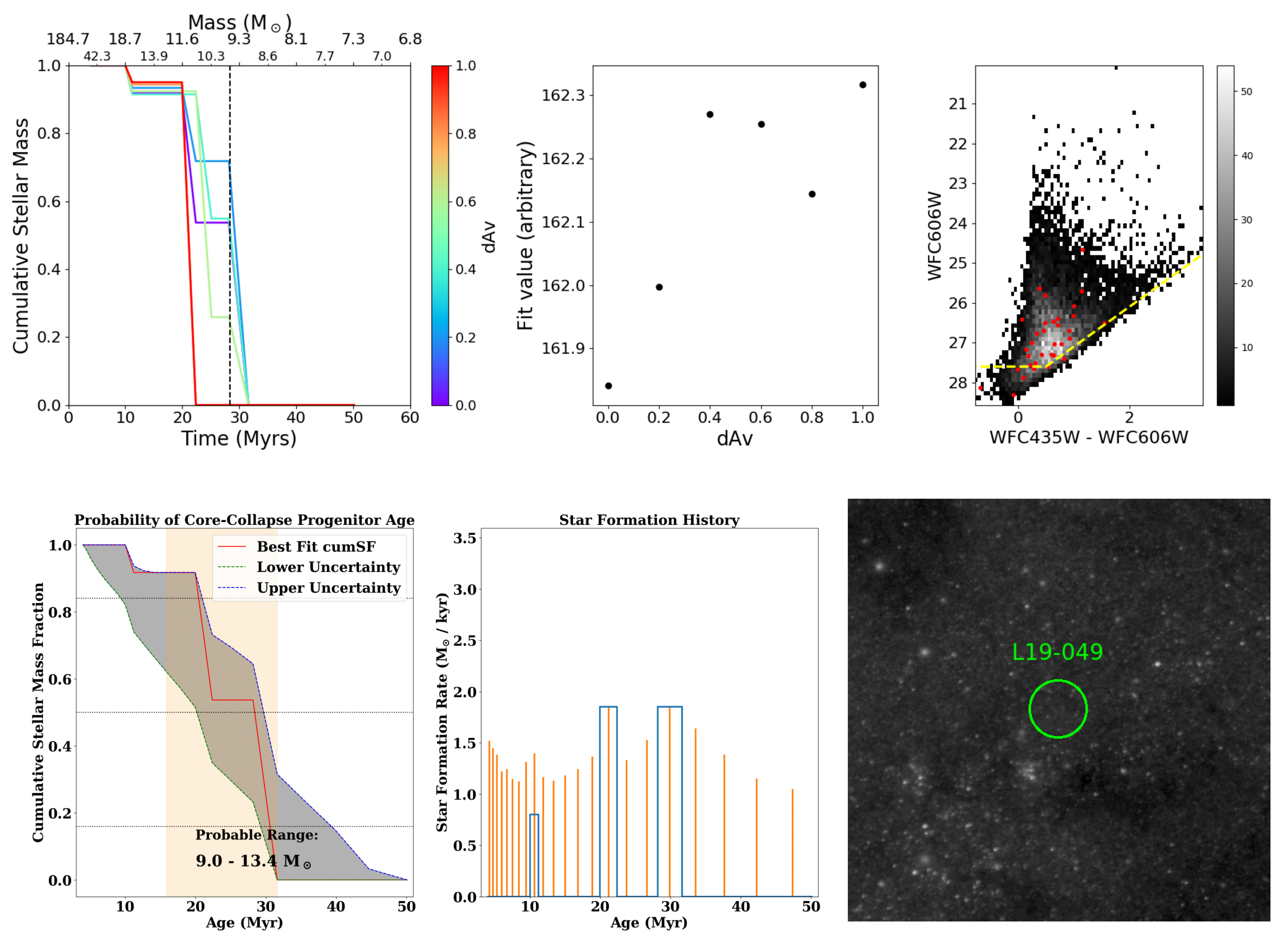}
     \caption{How we constrained SNR progenitor mass. \textit{Top left:} Multiple SFH measurements were made while assuming different amounts of $dA_{v}$ at the SNR location. Colored lines show the cumulative fraction of stellar mass in each age interval with the color denoting the amount of $dA_{v}$. \added{The vertical black dashed line indicates the SNR progenitor mass.} \textit{Top center:} Maximum likelihood fit values as a function of assumed $dA_{v}$. Lower values indicate a better fit. For L19-049, we report a $dA_{v}$ value of 0.0 as this produced the best fit to the data. \textit{Top right:} Observed CMD within our region of interest. The population in the background field are plotted in grey scale with darker regions indicating fewer stars. The population within 50 pc of the SNR are plotted as red points. The magnitude limits of the data are shown as \replaced{green}{yellow} dashed lines. These are based on the completeness for regions with the similar stellar density and photometric depth as determined by the Artificial Star Tests, see Section \ref{sect2.4.1} for details. \textit{Bottom left:} Examines the best SFH fit and the uncertainties produced by \texttt{hybridMC}. The red line shows the cumulative fraction of stellar mass in each age bin. The tan region indicates the probable mass range inferred from the range of ages consistent with the population median age. \added{The gray shaded region shows the 1$\sigma$ uncertainties on the most likely SFH.} \textit{Bottom center:} The differential SFH, showing the star formation rates and uncertainties that correspond to the cumulative fraction plot in \textit{bottom left}. The blue line indicates the most likely SFH for the region \added{, column 3 of Table \ref{tab4}. The orange lines show the maximum SF possible in each time bin, the sum of columns 3 and 5 from Table \ref{tab4}}. Bins with large uncertainties have a large covariance with neighboring bins which makes the cumulative distribution easier to interpret. \textit{Bottom right:} A 20" $\times$ 20" F606W image of the region of interest for the SNR with a green circle indicating the 50 pc extraction region for the resolved stellar photometry sample. \textit{We include figures of this format for all \replaced{175}{186} SNRs, 8 historically observed SNe, and the black hole formation candidate with progenitor constraints in the supplemental materials.}}
     \label{fig4}
 \end{figure}

\begin{figure}
     \centering
     \epsscale{1.2}
     \plotone{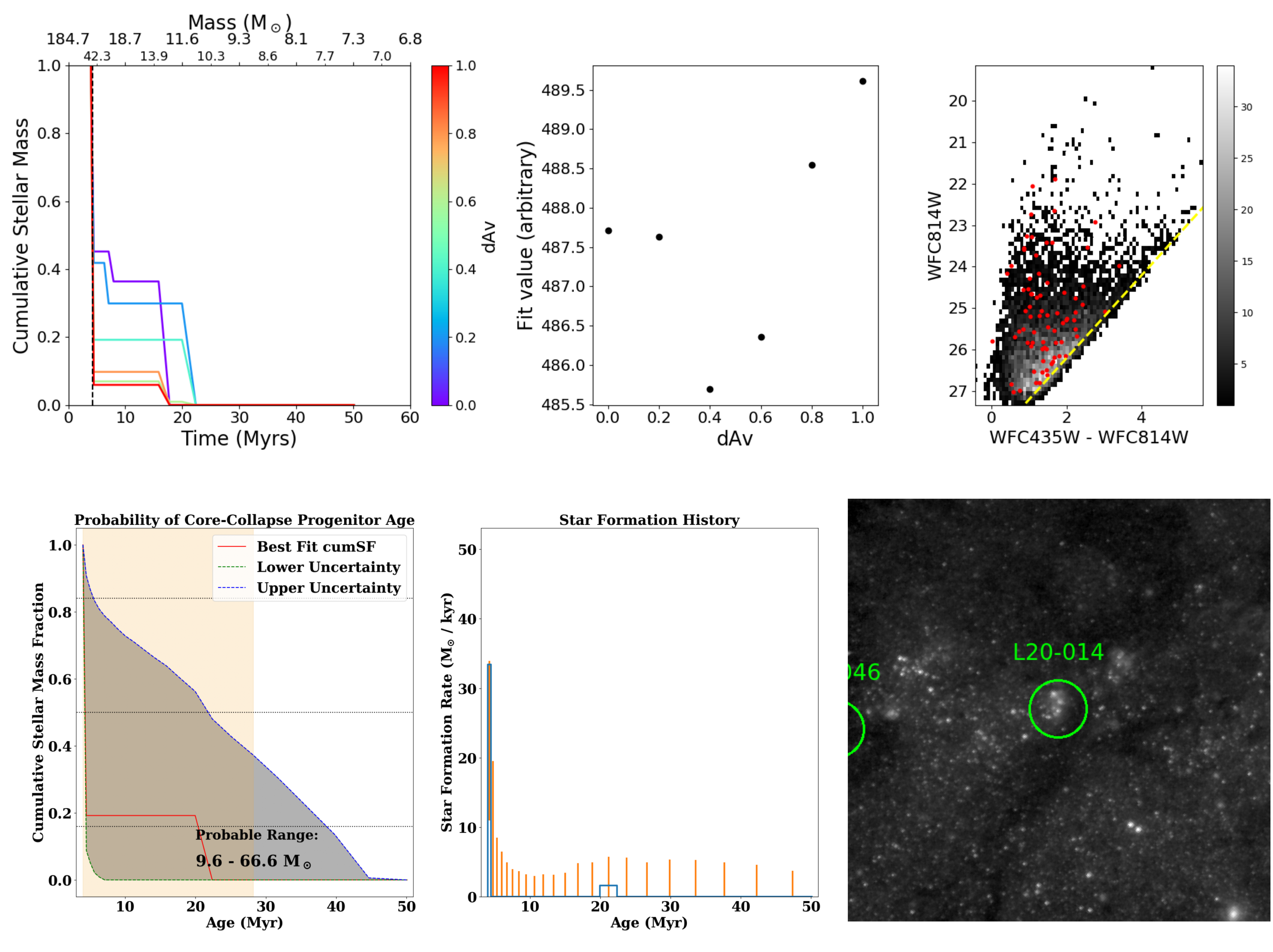}
     \caption{Same as Figure \ref{fig4}, but for L20-014. This shows an extreme case since L20-014 has one of the most massive inferred progenitors in our catalog.}
     \label{fig5}
 \end{figure}

\begin{figure}
     \centering
     \epsscale{1.1}
     \plotone{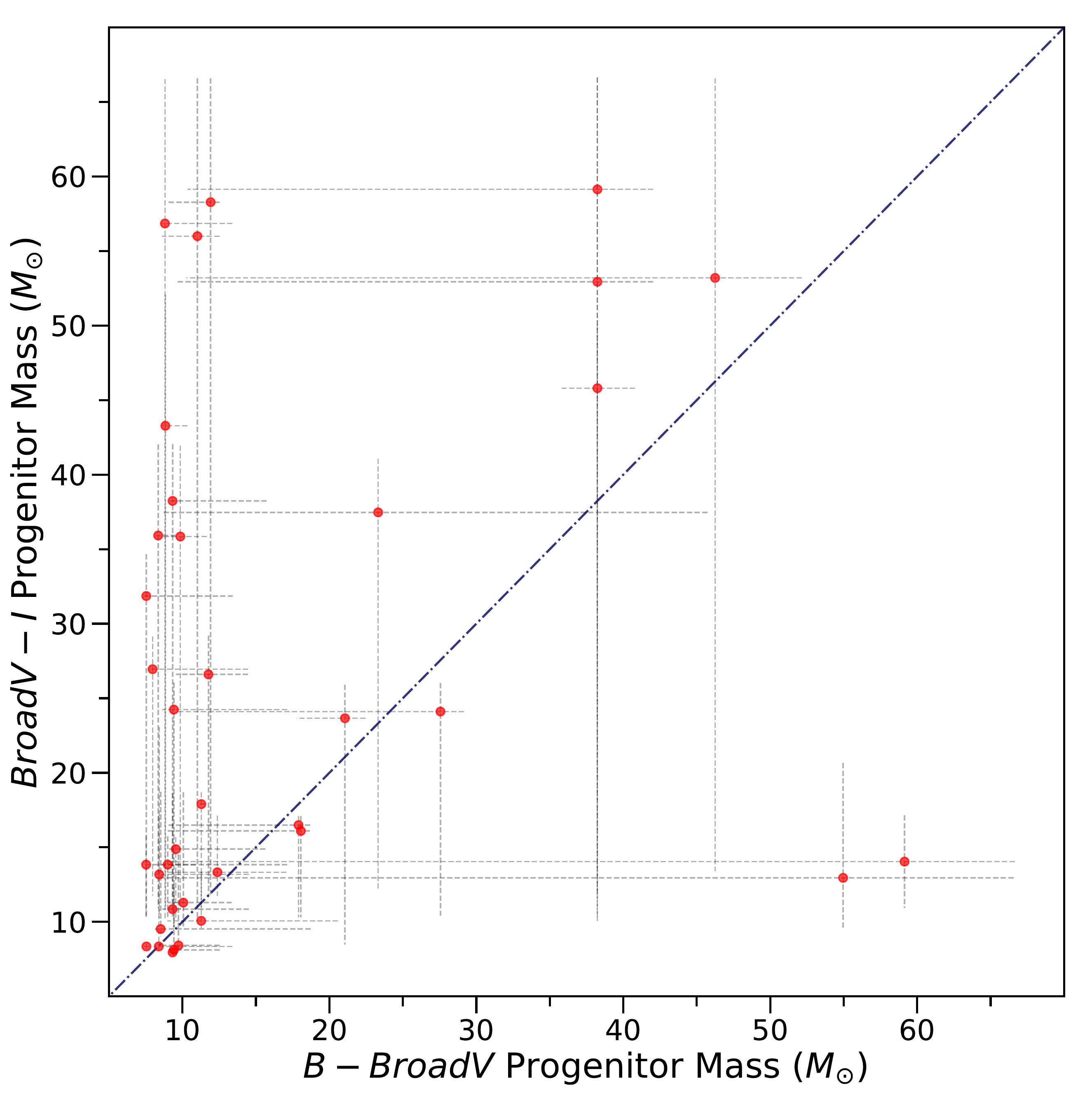}
     \caption{Comparing the inferred progenitor masses for the 37 locations we had data in three filters. The gray lines show the uncertainties associated with each mass. The dark blue line shows where the B$-$Broad-V progenitor mass is the same as the Broad-V$-$I mass. Few remnants fall along the equal mass line with most lying above it. This shows that using B$-$Broad-V CMDs biases our estimates towards lower masses while Broad-V$-$I CMDs biases towards higher masses. This is likely due to Broad-V band being affected by gas emission, see Section \ref{sect2.6}.}
     \label{fig6}
 \end{figure}

\begin{figure}
     \centering
     \epsscale{1.1}
     \plotone{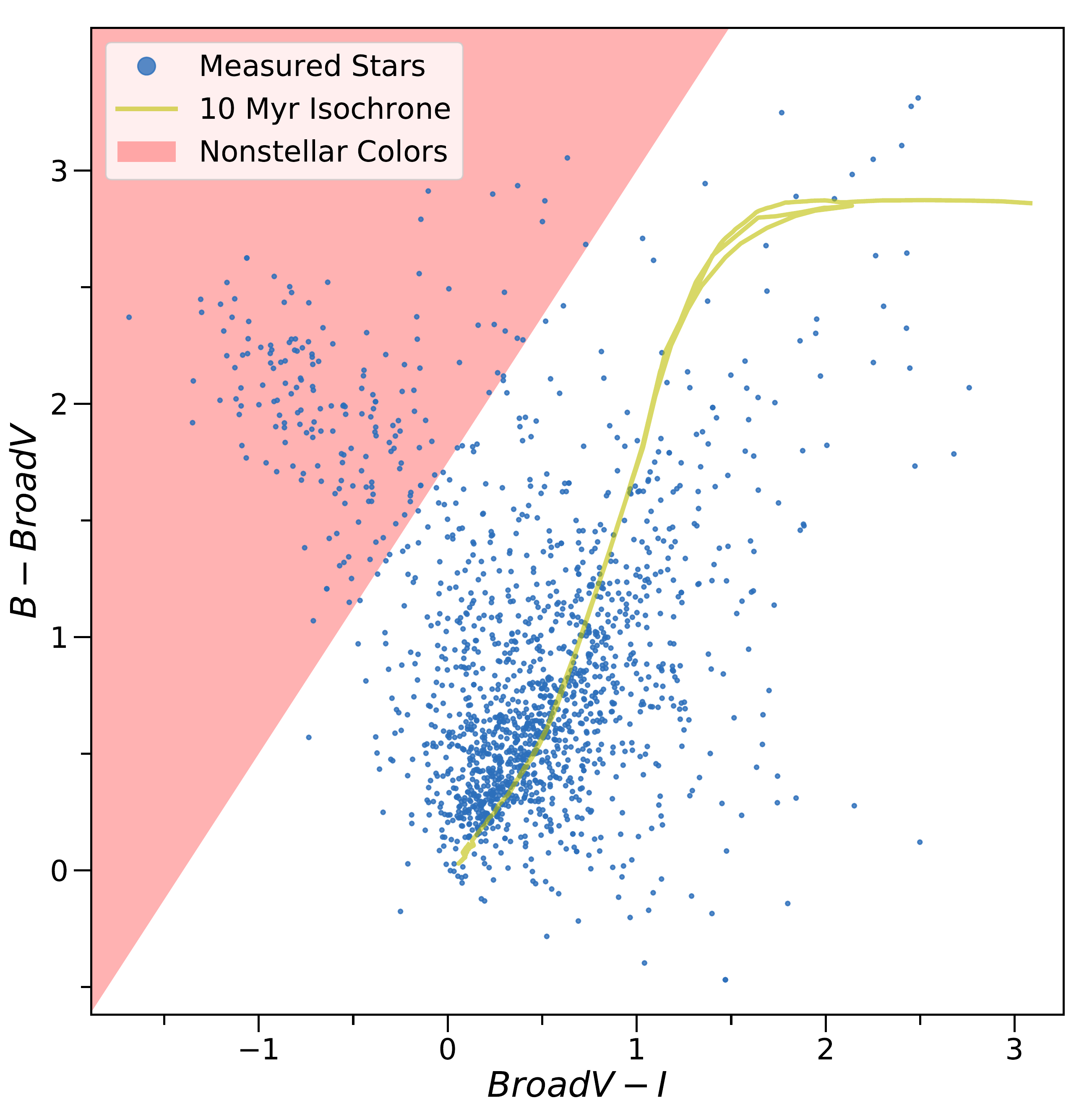}
     \caption{Broad-V$-$I color v. B$-$Broad-V color of the populations around the 37 remnants with data in three filters. The blue points indicate measured stars. The yellow line indicates where a 10 Myr old population is expected to reside in this parameter space \citep{Bressan12}. The red shaded region indicates the part of parameter space that is not consistent with the colors of stars. Dots in the region are likely gas emission or stars that have had their Broad-V flux overestimated, see Section \ref{sect2.6} for details.}
     \label{fig7}
 \end{figure}

Based on L19-098's line emission measurements, the Broad-V flux is $6.8\times10^{-14}$ ergs cm$^{-2}$ s$^{-1}$.
If we compare the Broad-V$-$I colors of the main sequence of this region to those typical for NGC 6946 for this color from Figure \ref{fig2}, we find that this gas emission caused a systematic Broad-V$-$I bias of $\sim$1.5 mag.
The B$-$Broad-V is redder than the main sequence in Figure \ref{fig2} by about the same amount resulting in a total difference of $\sim$3 magnitudes between B$-$Broad V and Broad-V$-$I.
Thus, it appears the gas emission of this strength can affect the Broad-V photometry by about 1.5 mag.
Most SNRs in the L19 catalog have gas emission line strengths that are $\lesssim$10\% that of L19-098, with $\sim$95\% having Broad-V fluxes less than $4\times10^{-15}$ ergs cm$^{-2}$ s$^{-1}$.
This suggests that for many SNRs the impact will be $\lesssim$0.15 mag.
That shift in color is similar to that seen in the other example SNR L20-023, shown in the right panel of Figure \ref{fig10}.
One cannot simply scale the gas impact from the SNR fluxes, since the gas emission can also originate from an associated H II region, which would not be included in the L19 SNR emission line flux measurements.

Such gas emission effects are not an issue for every remnant that uses Broad-V data, but in Figure \ref{fig6} we see that about one third of the measurements could be affected.
This fraction will negatively impact studies of the sample as a whole.
In particular, because the number of massive progenitors is small, a few spurious massive measurements present due to this effect and included in the sample could significantly bias the distribution.
Thus, while we leave measurements based on Broad-V data in our catalog, as most do not have large amounts of bright gas emission, we avoid using the Broad-V data in our samples when constraining the properties of the progenitor distribution.

We do not attempt to remove only locations affected by gas emission, because that could introduce a selection bias that would push our overall distribution towards older ages.
Younger populations typically reside in dustier regions of a galaxy so removing locations with lots of dust would get rid of those that are more likely to be very massive progenitors.
Additionally, young populations are found inside clouds of gas making them the most likely to have this contamination.

\added{In many cases DOLPHOT is able to handle this impact, however there are some locations that have a geometry that allows this to happen, i.e. when the gas emission is at the location of a star but not where DOLPHOT calculates the background sky. 
One could compare different DOLPHOT parameters to see if the impact could be minimized but this would likely require tailoring the parameters to each location and would not eliminate all issues with one set of locations.}

As a result of the potential bias in Broad-V photometry from gas emission at some SNR locations, we split our catalog into 5 subsamples.
The first consists of the SNRs that were fit using a B$-$Broad-V CMD.
The next subsample is the SNRs fit with a Broad-V$-$I CMD.
We also look at the SNRs with data in three filters.
We then consider the SNRs that didn't use Broad-V band when fitting (i.e. B$-$I and V$-$I CMDs).
The last consists of the historically observed events.
These locations did not show evidence for strong gas emission, and in the \replaced{6}{5} cases where we had three bands, our progenitor mass results were consistent across all band combinations.
In order to avoid any biases being introduced to the sample from a few masses being influenced by gas emission, we chose our preferred subsample to be the SNRs that did not use Broad-V band data to constrain the progenitor mass as well as the historically observed events.

\section{RESULTS} \label{sect3}

Our final measurements are compiled in Table \ref{tab2}.
Column 1 is the SNR identifier.
Column 2 identifies the bluer filter used.
Column 3 gives the 50\% completeness magnitude for the filter.
Column 4 identifies the redder filter used.
Column 5 gives the 50\% completeness magnitude for the filter.
The completeness magnitudes come from the artificial stars used while fitting, so locations with the same fake stars will have the same limits as determined from the assigned ASTs.
These were also the magnitude limits applied during the fitting.
Column 6 gives the number of stars in the region that were used to fit the SFH.
Columns 7 and 8 specify the best fitting differential and foreground reddenings applied to the region during the fitting process, respectively.
Column 9 has the inferred progenitor mass from the median age as well as bounds in M$_{\odot}$.
Missing data in column 9 indicate locations where either no SF was measured in the past 50 Myr or we were unable to measure the SFH.

We present an example age probability distribution for SNR L19-049 in Table \ref{tab4}.
Similar tables for all \deleted{measured} locations \added{were the best fit SFH contained SF within the last 50 Myr} are available in the supplemental materials.
Columns 1 and 2, T1 and T2, define the beginning and end (respectively) of the age bin in $log(age)$.
Column 3, SFR(best), is the best-fit star formation rate, in that age bin in $M_{\odot}/yr$ \added{and is plotted as the blue line in Figure \ref{fig4}}.
Columns 4 and 5 are the negative and positive error bars on that rate \added{, plotted as orange lines in Figure \ref{fig4}}.
Column 6, PDF(best), shows the fraction of stellar mass $<$50 Myr in the age bin according to the best-fit SFH equivalent to the probability of that progenitor age.
Columns 7 and 8 are the negative and positive error bars on that probability.
Column 9, CDF(best), provides the cumulative fraction of stellar mass over time for the best-fit SFH.
\replaced{Columns 10, 11, and 12 are the 16th, 50th, and 84th percentiles}{Columns 10 and 11 are the lower and upper uncertainties} of the cumulative mass fraction for a set of a million realizations of the SFH with the uncertainties in columns 4 and 5.
Columns \replaced{13 and 14}{12 and 13}, M1 and M2, gives the mass of stars with lifetimes within the age bin in M$_{\odot}$.

Figure \ref{fig11} is a histogram of the progenitor masses in our sample of NGC 6946 and also the distributions from M31+M33 \citep{Diaz18} and M83 \citep{Williams19}.
Compared to these we inferred more progenitor masses between 20 and 40 M$_{\odot}$.
This is likely due to the gas contamination discussed in Section \ref{sect2.6}.

We were unable to determine the progenitor mass for 16 SNRs because there was no young population in the regions which indicate likely locations of Type Ia SNe.
As a Sbc type galaxy, NGC 6946 is expected to have $\sim$16.5\% of its SNe be Type Ia, however only 7.1\% of our sample are good Type Ia candidates, suggesting that at most 10\% of our progenitor masses may be false detections of Type Ia progenitors that happen to have occurred in regions with a young population present.
More of these would likely be consistent with no recent SF if uncontaminated data was used instead.

\subsection{Historical Supernovae} \label{sect3.1}

Within NGC 6946 there have been ten SNe observed and one star that may have collapsed directly into a black hole (NGC6946-BH1).
Eight of the historically observed SNe are known to be Type II core-collapse events but SN1939C is only classified as a Type I and SN1969P is without a claimed type \citep{Guillochon17}.
We were able to constrain the progenitor of eight of the observed SNe and NGC6946-BH1.
We did not recover enough stars around SN2004et for a fit to converge and thus are not able to constrain the progenitor.
The ranked progenitor mass distribution of the locations with mass estimates were best fit by a power-law function with an index of $-3.8^{+1.4}_{-2.2}$.
The small sample size is the source of our large errors.
As with most of the SNRs in our catalog, the historical SNe tend to reside in the spiral arms of the galaxy.
Most of these progenitors were constrained by \cite{Williams14a}, \cite{Williams18}, and \cite{Murphy18}.
However \cite{Williams14a} and \cite{Williams18} used a closer distance to NGC 6946 than us and \cite{Murphy18} used a larger radius for their region than we do.

In the following subsections we discuss the results of the ten events we constrained and compare to previous estimates.
The locations of all SNe before 2013 come from \cite{Lennarz12}.
While the location for SN2017eaw and NGC6946-BH1 come from \cite{Wiggins17} and \cite{Gerke15}, respectively.

\subsubsection{SN1917A (Type II)}

We inferred a mass of $9.3^{+16.7}_{-0.7}$ M$_{\odot}$ for SN1917A.
This is consistent with the estimates of \cite{Williams14a} ($7.9^{+6.7}_{-0.5}$ M$_{\odot}$) and \cite{Williams18} ($15.0^{+1.0}_{-5.0}$).

\subsubsection{SN1939C (Type I)}

SN1939C is the only observed Type I  \replaced{SNe}{CCSN} in NGC 6946 and has not had its subclass identified.
Our estimate of the progenitor mass of $31.9^{+2.8}_{-22.9}$ M$_{\odot}$ is the first for this SN.
Being high mass suggests a subclass of b or c is appropriate as their progenitors are thought to be massive Wolf-Rayet stars \citep{Smartt15}.
The age clearly indicated a core-collapse origin; however, the uncertainties in age allow for a possible $\sim$10 M$_{\odot}$ or more massive binary progenitor \citep{Yoon10}.

\subsubsection{SN1948B (Type IIP)}

We estimate the progenitor of SN1948B to be $8.9^{+5.6}_{-0.3}$ M$_{\odot}$.
This estimate is consistent with the previous estimate from \cite{Williams18} of $10.0^{+4.0}_{-1.0}$ M$_{\odot}$.

\subsubsection{SN1968D (Type II)}

We inferred a progenitor mass of $19.6^{+1.0}_{-11.0}$ M$_{\odot}$ for SN1968D.
This mass is higher than found by \cite{Williams18} \added{($9.6^{+4.2}_{-0.6}$ M$_{\odot}$)} but consistent within the errors.

\subsubsection{SN1969P (Unknown)}

We estimate the progenitor mass of SN1969P to be $8.8^{+4.6}_{-0.2}$ M$_{\odot}$ \added{, assuming it originated from a core-collapse event}.
This is the first time the mass of the progenitor has been constrained.
It is the only observed SNe in NGC 6946 that has not has its classification determined.
\deleted{The inferred mass points to it being a core-collapse event, most likely Type II.}

\subsubsection{SN1980K (Type IIL)}

Our estimate of the progenitor mass of SN1980K ($7.5^{+8.2}_{-0.2}$ M$_{\odot}$) is consistent with that of \cite{Williams18} ($9.0^{+6.7}_{-0.2}$ M$_{\odot}$).
They did not detect any young population near the SN and had to extend their radius out to 100 pc for their estimate.
Using a larger radius increases the probability of including stars which are not related to the SN progenitor when fitting the CMD.
This leads us to prefer our estimate as we did not extend our radius.

\subsubsection{SN2002hh (Type IIP)}

Previous CMD fitting resulted in an inferred progenitor mass of $15.9^{+4.8}_{-1.3}$ M$_{\odot}$ \citep{Williams14a}.
Here we infer a mass of $8.4^{+2.5}_{-0.3}$ M$_{\odot}$.
\cite{Williams14a} did not use a contamination CMD when fitting, which is likely the cause of the difference.
As discussed in Section \ref{sect2.4.4}, not including a contamination CMD results in less statistically significant age estimates.
This leads us to prefer our estimate to the previous estimate.

\subsubsection{SN2008S (Type IIn-pec)}

SN2008S was identified as a possible SN imposter by \cite{Steele08} based on its spectral properties and faint magnitude.
SN imposters are thought to be superoutbursts of massive luminous blue variable stars.
\cite{Prieto08} found the likely progenitor of SN2008S in $Spitzer$ mid-IR images but was unable to find it in optical images.
Follow up images show this progenitor continues to fade in IR bands while remaining elusive in the optical \citep{Prieto10,Adams16}.
Using follow up spectroscopy and photometry \cite{Botticella09} concluded that the observed properties of the progenitor could be explained by a electron capture supernovae \citep{Tominaga13}.
These occur when a super-asymptotic giant branch star with a main-sequence mass of $\sim$7-9.5 M$_{\odot}$ become a CCSN.
We infer a mass of $8.0^{+9.1}_{-0.3}$ M$_{\odot}$, which is consistent with the $7.5^{+0.2}_{-0.2}$ M$_{\odot}$ from \cite{Williams18}.
These match theoretical models of electron capture SNe suggesting that SN2008S may not be an imposter but an uncommon type of CCSNe.

\subsubsection{SN2017eaw (Type IIP)}

Precursor images of the stars near the site of SN2017eaw has led to progenitor mass estimates between 13 and 15 M$_{\odot}$ \citep{KF18,VD19}.
However, other than the star that is thought to be the progenitor, there were no luminous stars within about 3" of the SN site \cite{VDe17}.
\cite{Williams18} used CMD fitting of the stellar population near the SN cite after explosion and inferred a mass of  $8.8^{+2.0}_{-0.2}$ M$_{\odot}$.
While our overall age probability distribution, and associated uncertainties, is similar to \cite{Williams18}, they measured a spike of SF around 30 Myr ago, which we did not.
\replaced{Our best fit did not contain any SF within the last 50 Myr, which is consistent with the uncertainties from \cite{Williams18}.}{The uncertainties associate with this spike are consistent with there being no SF in the time bin, which is what our best fit SFH contained.}
\added{Other than this, the best fit from \cite{Williams18} found no SF within the last 50 Myr making our measurements consistent with one another.}

\subsubsection{NGC6946-BH1}

NGC6946-BH1 is thought to be the first star that has been seen collapsing directly into a black hole.
Fitting the spectral energy distribution led to a progenitor age estimate of $\sim$25 M$_{\odot}$ \citep{Adams17}.
\cite{Murphy18} inferred a mass of $17.9^{+29.9}_{-7.6}$ M$_{\odot}$ using an age dating technique similar to ours.
Our inferred progenitor mass of $16.4^{+0.7}_{-7.4}$ M$_{\odot}$ is consistent with the previous population dating technique though not the spectral energy distribution estimate.
\cite{Murphy18} suggests that the difference in estimates is due to the uncertainty in mapping late evolutionary stage energy distributions to progenitor masses as well as no stars as bright as the progenitor were found within 300 pc.

\section{DISCUSSION} \label{sect4}

With the mass constraints we are able to learn how SN activity is distributed within NGC 6946 and constrain the progenitor mass function index.
To investigate the spatial distribution of SNe, in Figure \ref{fig12} we plot the locations of our catalog on a V band image taken with Kitt Peak's 2.1m telescope where mass is indicated by color.
\added{Our progenitor mass constraints also allowed us to investigate the distribution of initial stellar masses that produce supernovae in NGC 6946.
We described each of these in turn below.}
\deleted{To constrain the progenitor mass function index, we use Kolmogorov-Smirnov (KS) tests to determine the likelihood each subsample, ranked by mass, is drawn from various power-law distributions.
KS tests return a p-value which indicates how likely it is the subsample is drawn from the same parent distribution.
A p-value of 1 means there is a 100\% chance while a p-value of 0 means there is a 0\% chance.
Here we report the indices in which the p-values were $\geq$0.1, highlighting the most likely index as well.
Future work will constrain the progenitor mass function index using a more robust Bayesian technique to properly account for the uncertainties.}

\subsection{Spatial Distribution of Progenitor Masses} \label{sect4.1}

The inferred progenitor masses are plotted on an image of NGC 6946 in Figure \ref{fig12} to show the distribution across the entire galaxy.
The type and mass of the progenitors are indicated by the symbol and color, respectively, used in the image.
Locations in our catalog that we were able to measure SF within the last 50 Myr (flags ``p" and ``h") are shown as colored circles.
Progenitors with masses above 20 M$_{\odot}$ are red, masses between 16 and 20 M$_{\odot}$ are orange, masses between 12 and 16 M$_{\odot}$ are yellow, and masses less than 12 M$_{\odot}$ are blue.
We indicate historically observed events by \deleted{larger} colored \replaced{circles}{diamonds}.
The measurements that do not use the Broad-V band (flag ``p" only) are shown as thicker circles since these are the locations that are less likely to be biased by gas emission (see Section \ref{sect2.6}).
Locations with no young population, i.e. good Type Ia SNe candidates, are shown as \replaced{green}{yellow} squares (flag ``n").
The SNRs mostly trace out the spiral arms of the galaxy but we do not see any other significant clumping even when comparing our subsamples.

\subsection{Progenitor Mass Distribution} \label{sect4.2}

\added{Determining the masses of SN progenitors, and how they are distributed, allows us to better constrain the masses at which stars collapse directly into black holes as well as provide incites into the Red Supergiant Problem.
There are sophisticated Bayesian techniques for constraining the progenitor mass \citep{Diaz21}, which will be done by future works to properly account for the uncertainties.
In order to get a rough approximation of how our results compare to other works, we performed a simplified method for estimating the progenitor mass distribution.

We use Kolmogorov-Smirnov (KS) tests to determine the likelihood each subsample, ranked by mass (column 9 of Table \ref{tab2}), is drawn from various power-law distributions.
KS tests return a p-value which indicates how likely it is the subsample is drawn from the same parent distribution.
A p-value of 1 means there is a 100\% chance while a p-value of 0 means there is a 0\% chance.
Here we report the indices in which the p-values were $\geq$0.1, highlighting the most likely index as well.}
\deleted{To determine the progenitor mass distribution index of each sample we ran a series of KS tests.
We tested the ranked list of masses from column 9 of Table \ref{tab2} against a series of power-law functions, finding the power-laws that were most likely drawn from the same parent distribution as the measured masses.
These tests gave us the range of indices in which there is a $\geq$90\% chance that they come from the same parent distribution along with the index that is most likely.}

We fit our full catalog to power-law indices and determine it was best fit by an index of -2.4 with a 90\% confidence range of -2.5 and -2.3.
A Salpeter index (-2.35; \cite{Salpeter55}) is included in the errors but this is likely a result of the B$-$Broad-V bias working to cancel the effects of the Broad-V$-$I bias.
As seen in Figure \ref{fig11}, more masses than expected were inferred between 20 and 40 M$_{\odot}$ as well as above 55 M$_{\odot}$ as compared to \cite{Diaz18} and \cite{Williams19}.
These extra SNRs with high inferred progenitor masses caused our KS tests to return lower p-values overall, with a peak at 0.21.

\begin{figure}
     \centering
     \includegraphics[width=\linewidth]{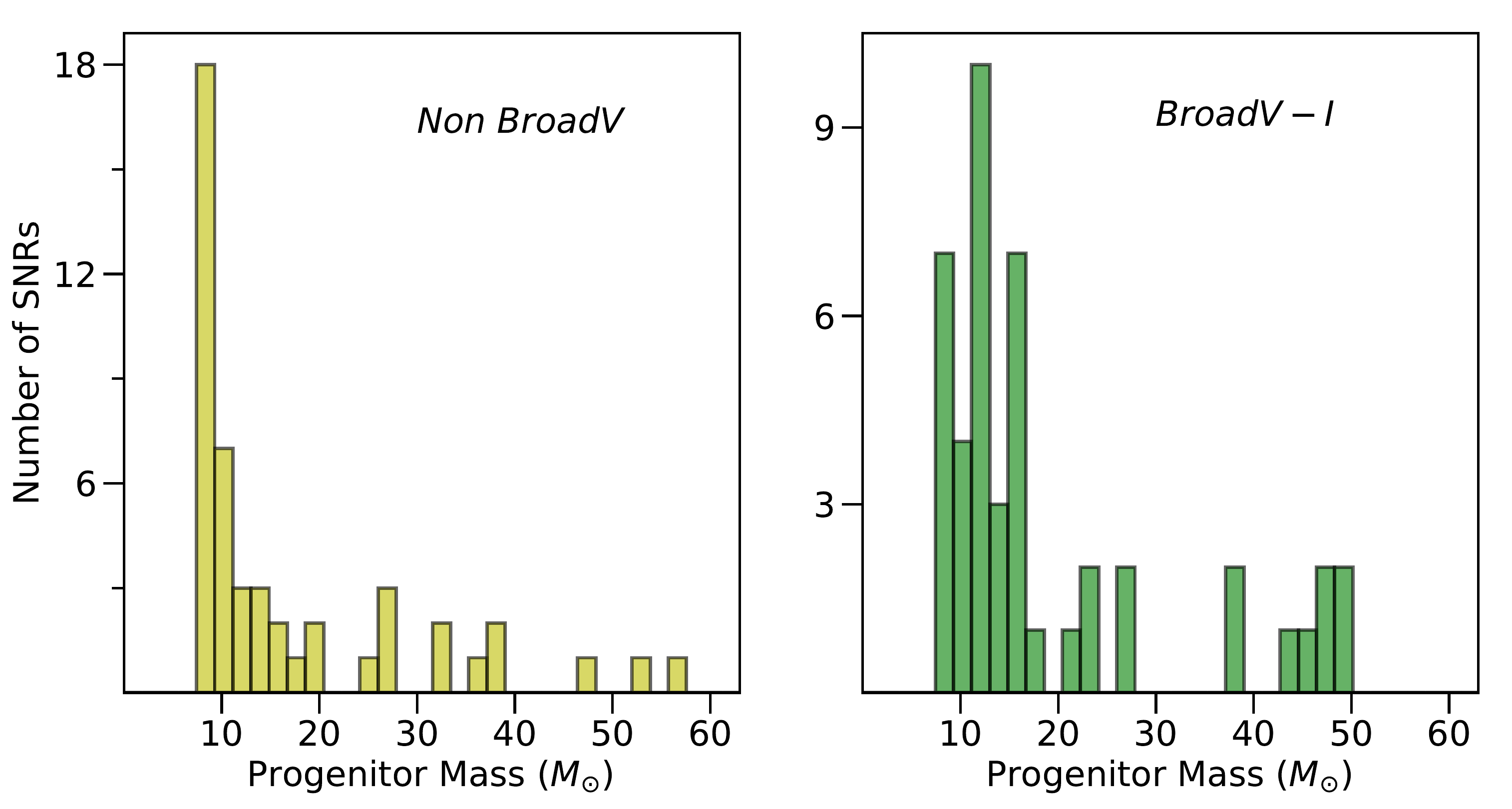}
     \includegraphics[width=0.5\linewidth]{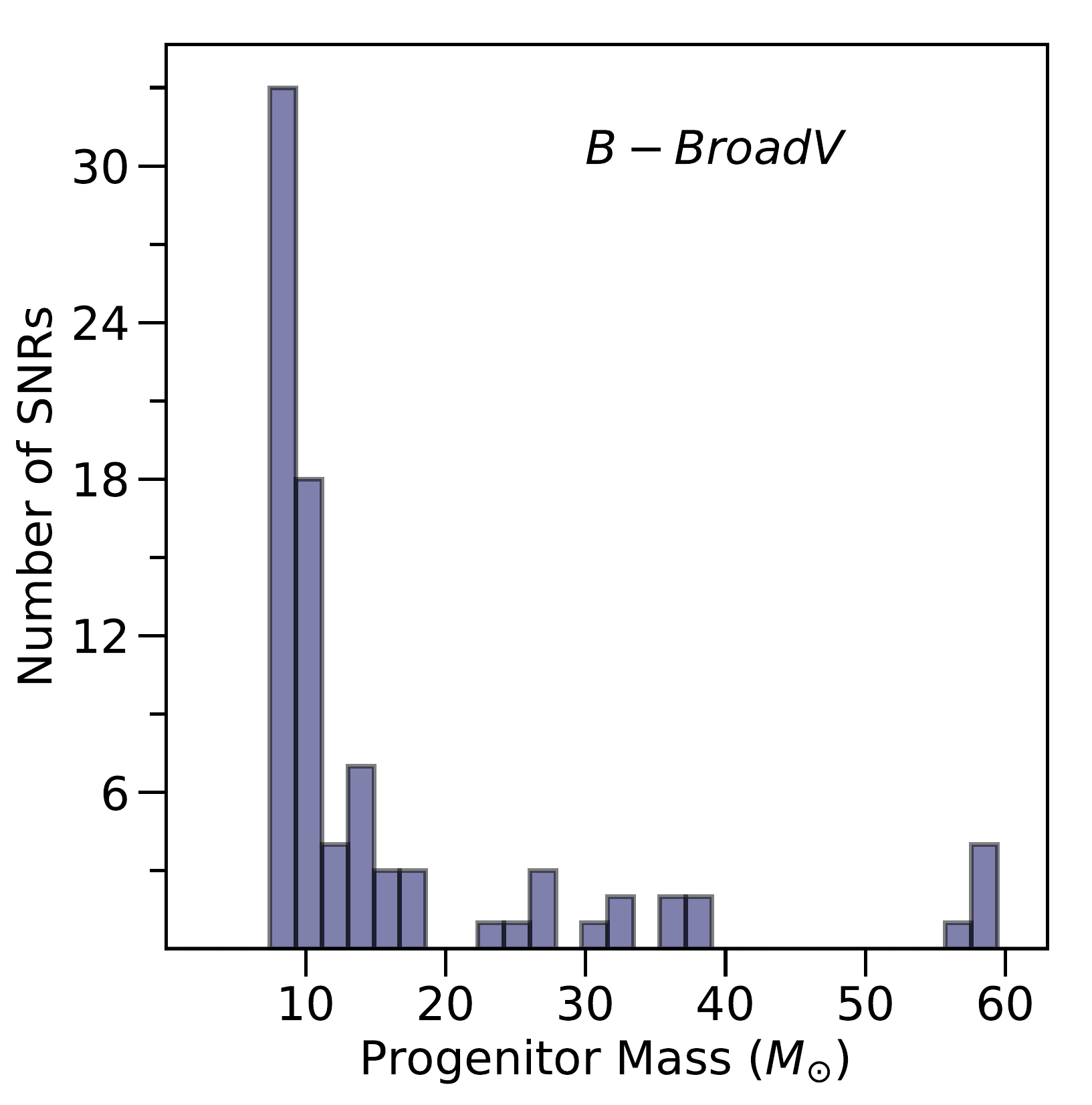}
     \caption{A comparison of inferred mass distributions using different filter combinations. \textit{Top left:} Distribution of masses inferred using CMDs that did not use Broad-V data. \textit{Top right:} Distribution of masses inferred from Broad-V$-$I CMDs. \textit{Bottom:} Distribution of masses inferred from B$-$Broad-V CMDs. The Broad-V$-$I CMDs resulted in a flatter distribution than the other two. The B$-$Broad-V distribution has no masses between 40 and 55 M$_{\odot}$. Both are expected given how gas emission affects the inferred mass, see Sections \ref{sect2.6} and \ref{sect4.2} for details.}
     \label{fig8}
 \end{figure}

\begin{figure}
     \centering
     \epsscale{1.2}
     \plotone{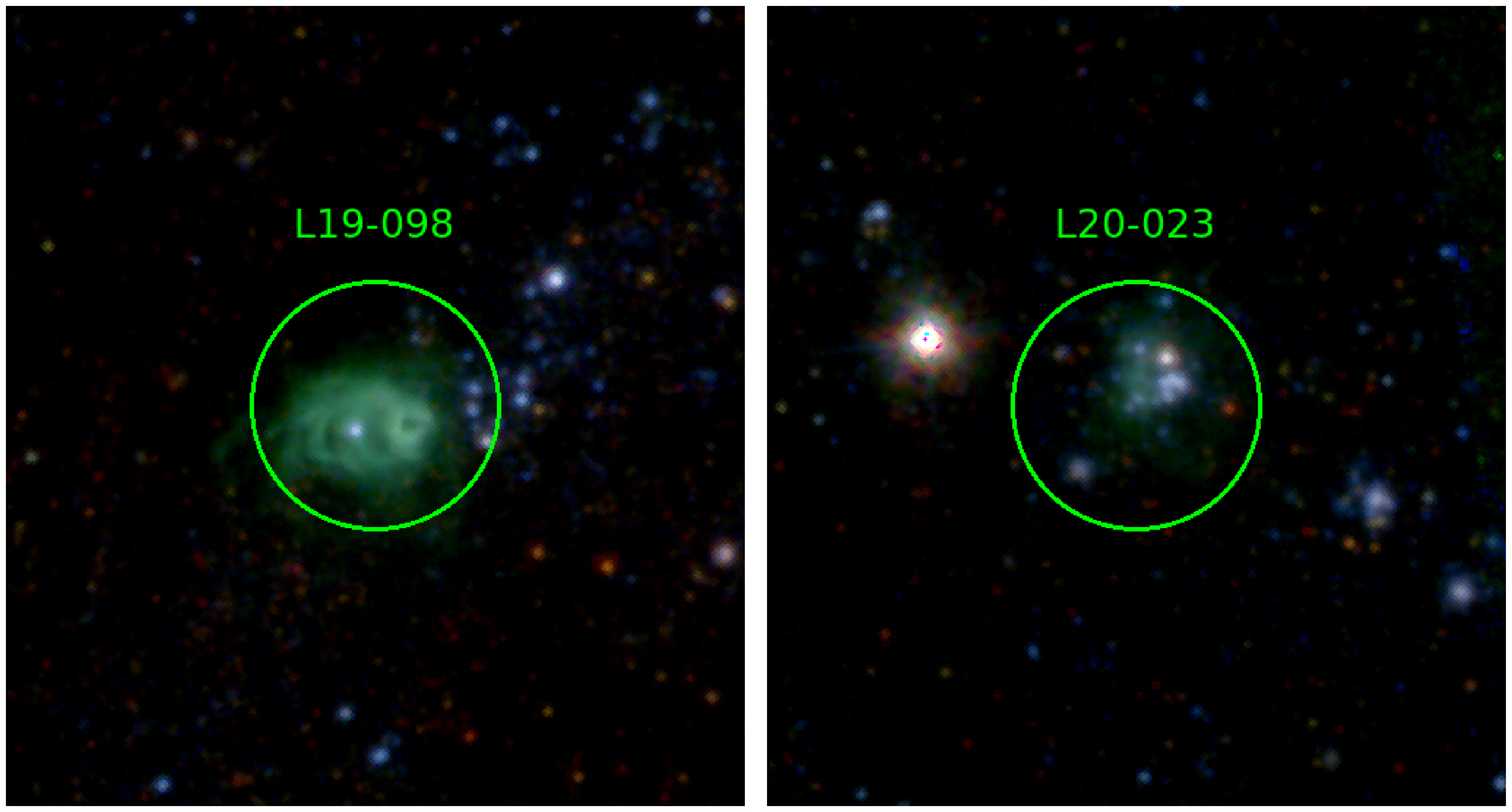}
     \caption{Illustrating the impact nebulae and surrounding gas emission have on mass estimates. Each is an 8" x 8" colored image with B band data in blue, Broad-V band data in green, and I band data in red. The circles show the 50 pc regions used in the fitting process. The right image shows L20-023 while the left shows L19-098. L20-023 is a more typical SNR while L19-098, as a known ultraluminous x-ray source, is an extreme example. Both SNRs had a large difference in the inferred progenitor mass based on which CMD was used. L19-098's progenitor was inferred to be 8.0 M$_{\odot}$ when a B$-$I CMD was used, 8.4 M$_{\odot}$ with a B$-$Broad-V CMD, and 35.9 M$_{\odot}$ with a Broad-V$-$I CMD. L20-023 was inferred to be 27.4 M$_{\odot}$, 8.8 M$_{\odot}$, and 43.3 M$_{\odot}$ respectively. For a full discussion on the impact gas emission has on our inferred progenitor masses see Section \ref{sect2.6}.}
     \label{fig9}
 \end{figure}

\begin{figure}
     \centering
     \epsscale{1.2}
     \plotone{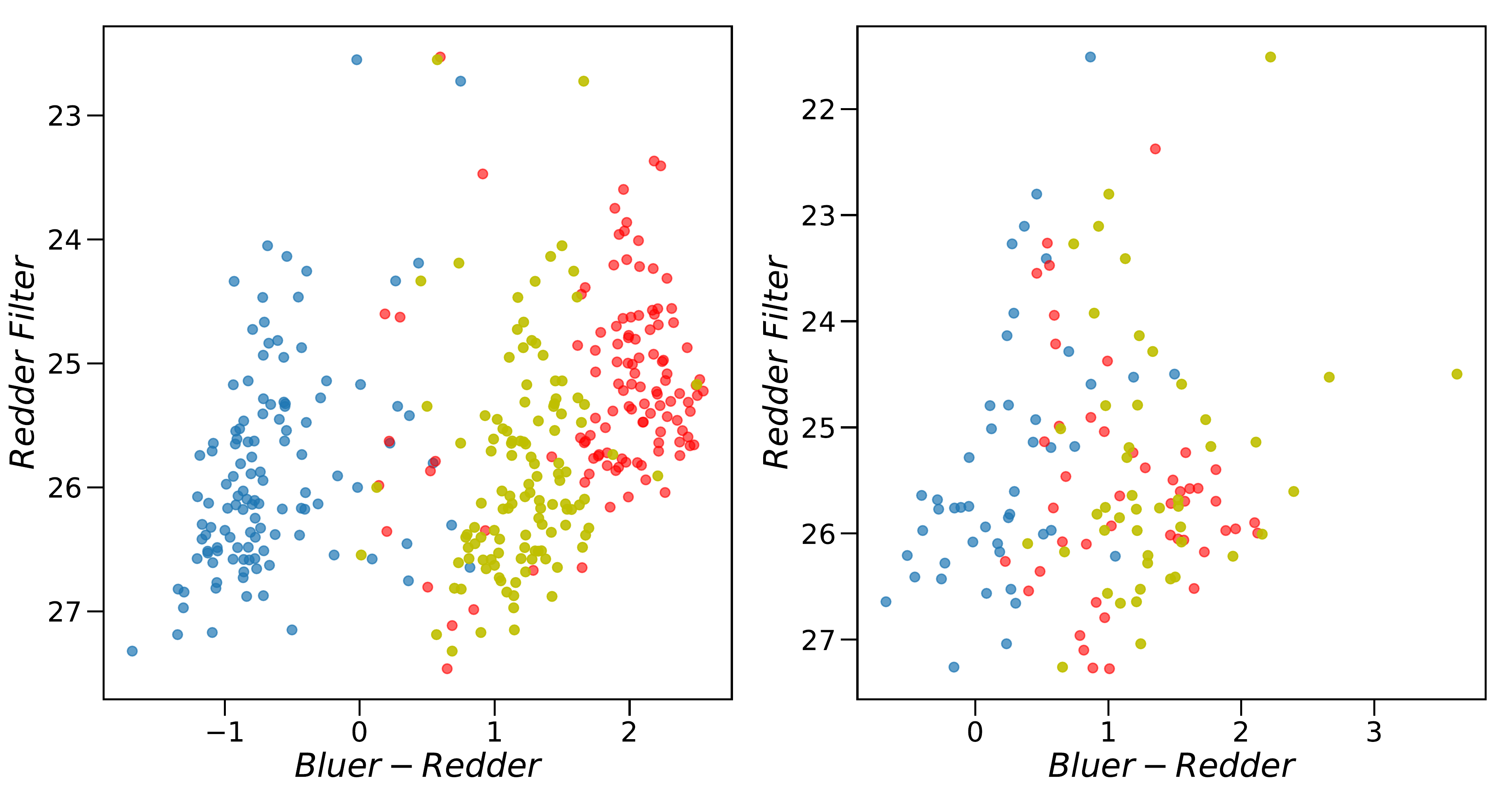}
     \caption{A combined CMD showing the biases in using Broad-V band. Blue points show where the stars lie in a Broad-V$-$I CMD, the red for B$-$Broad-V, and yellow for B$-$I. The stars in the right plot are near L20-023 that passed our quality cuts in B, Broad-V, and I bands, while the stars in the left plot are those that passed our cuts near L19-098. In both examples the Broad-V$-$I points are too blue making our age estimates younger. For L19-098, the B$-$Broad-V points are too red making the age estimate older. For a full discussion on the impact of gas emission see Section \ref{sect2.6}.}
     \label{fig10}
 \end{figure}

\begin{figure}
     \centering
     \epsscale{1.1}
     \plotone{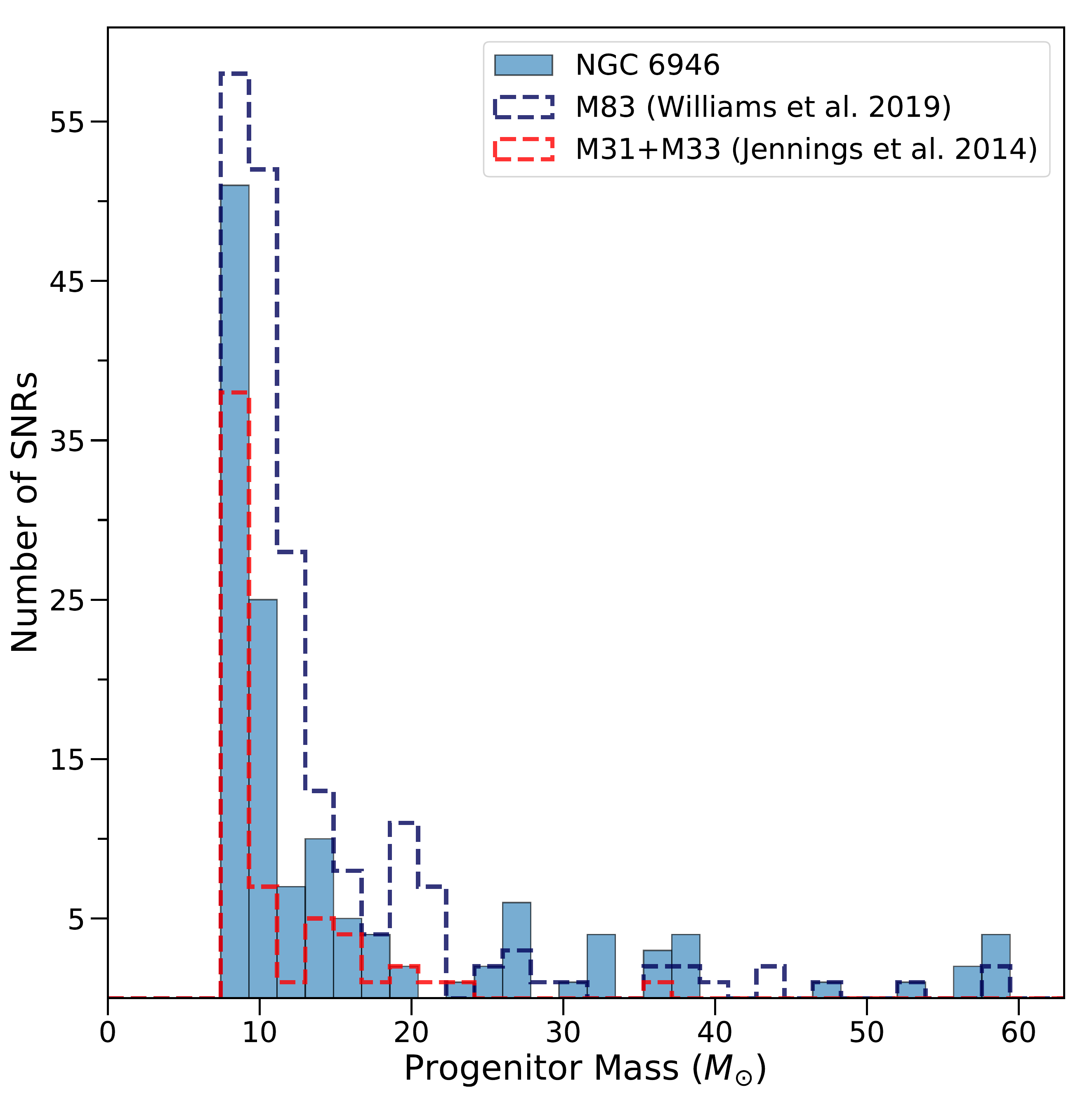}
     \caption{The teal histogram shows the NGC 6946 distribution of progenitor masses that were measured using a CMD other than Broad-V$-$I, with the M83 (dark blue dashed line) and M31+M33 (red dashed line) distributions shown for comparison. NGC 6946 shows more progenitors with mass $>$20 M$_{\odot}$ than M31+M33, though we didn't recover as many around 20 M$_{\odot}$ as in M83. The cutoff at lower masses was measured for the M31+M33 sample \citep{Diaz18}, but was assumed for this work, see Section \ref{sect2.5} for details.}
     \label{fig11}
 \end{figure}

\begin{figure}
     \centering
     \epsscale{1.18}
     \plotone{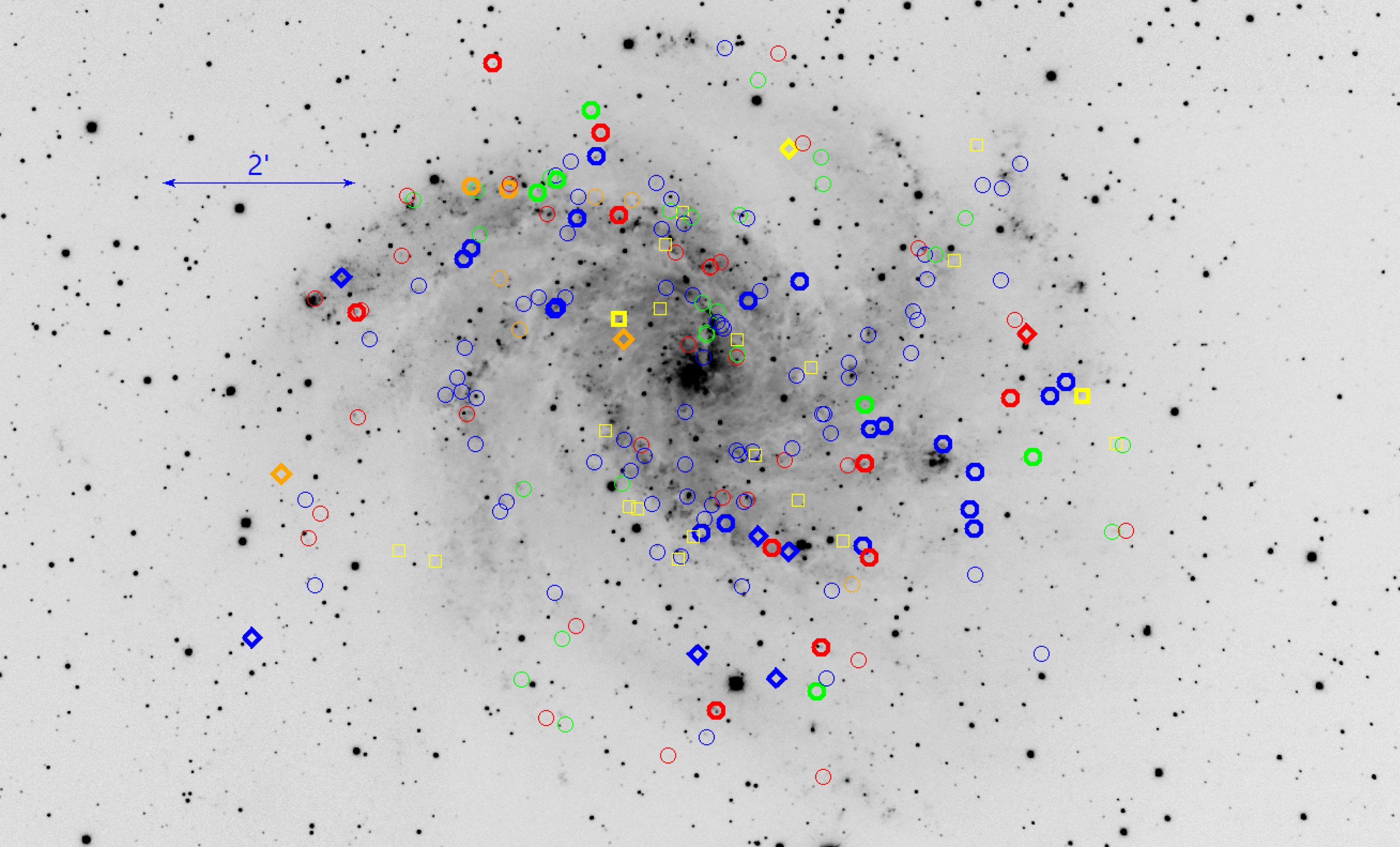}
     \caption{Locations of NGC 6946 SNRs color-coded by the progenitor masses inferred from their local stellar population are overplotted on a V band image from Kitt Peak National Observatory's 2.1m telescope. Colored circles indicate: red for $>$20 M$_{\odot}$; orange for 16-20 M$_{\odot}$; yellow for 12-16 M$_{\odot}$; blue for $<$12 M$_{\odot}$. SNRs with no local young population (Type Ia candidates) are plotted as yellow squares. The historically observed events that we were able to determine the progenitor mass for are indicated as \replaced{lightly larger colored circles}{diamonds}. \added{Yellow diamonds indicate historical SNe where the best fit SFH did not contain any SF in the past 50 Myr.} Locations where \replaced{H$\alpha$}{gas} contamination is not an issue are indicated by thicker \replaced{circles}{lines}.}
     \label{fig12}
 \end{figure}

To further quantify the effects of gas emission discussed in Section \ref{sect2.6} on our inferred progenitor masses, we fit the B$-$Broad-V, Broad-V$-$I, and non-Broad-V subsamples to power-law distributions.
Figure \ref{fig8} shows histograms comparing these distributions.
The B$-$Broad-V distribution was best fit by a power-law index of -2.9 with a 90\% confidence range between -3.4 and -2.8.
This is expected given how many low mass progenitors were inferred.
The Broad-V$-$I distribution resulted in a best fitting index of -1.8 (90\% confidence between -2.1 and -1.5), as expected given how evenly spread the masses are in Figure \ref{fig8}.
These are not consistent with one another, suggesting that the bias introduced by the Broad-V filters is strong enough that it cannot be reliably included in a statistical analyses of the sample.
However, we note that individual measurements may still be reliable in cases where there is not significant gas emission present.

\begin{figure}
     \centering
     \epsscale{1.2}
     \plotone{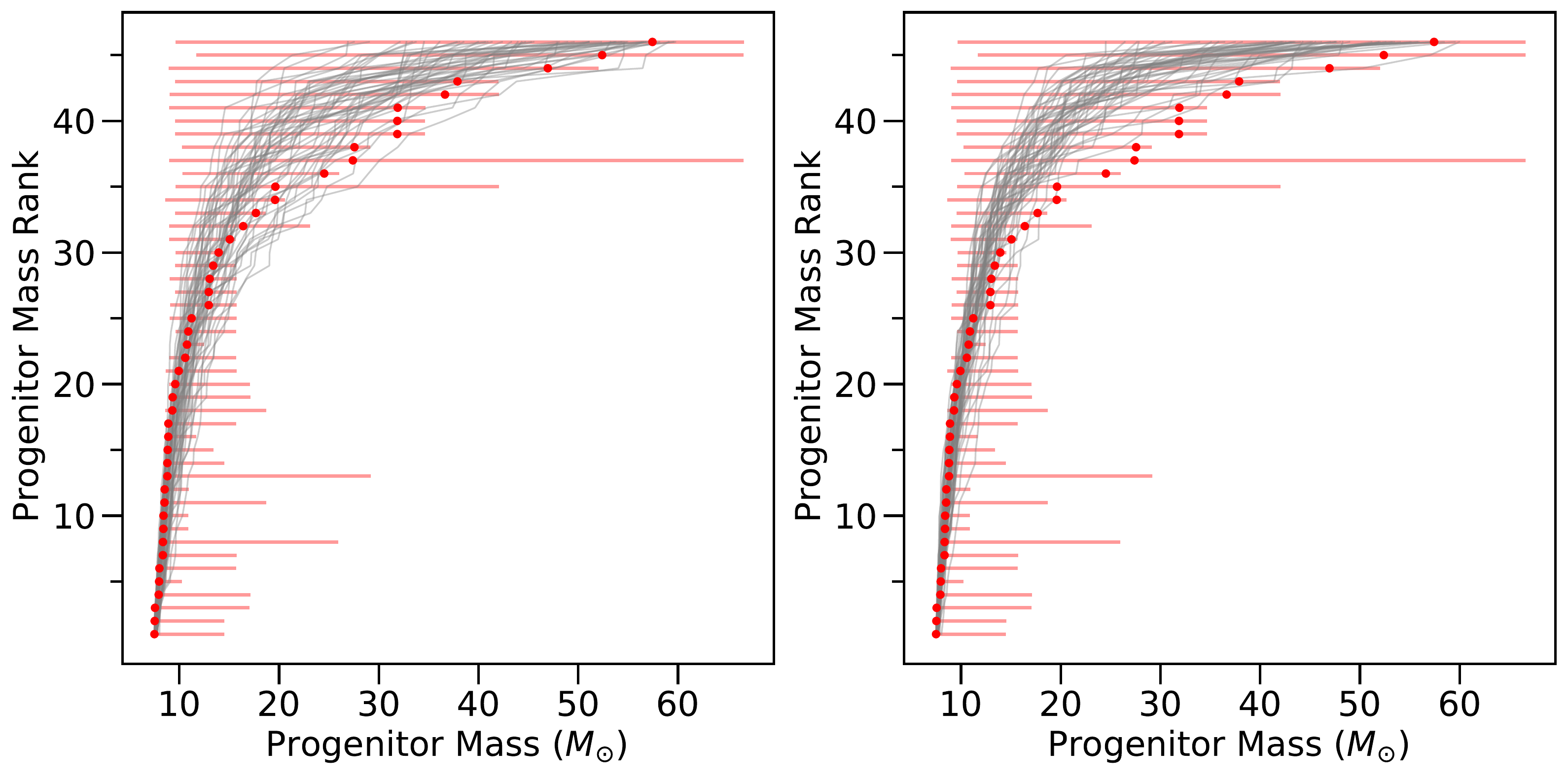}
     \caption{The ranked distribution of progenitor masses for the subsample of our catalog with reliable estimates, see Section \ref{sect4.2} for details.
     Overplotted are 50 gray lines showing draws from a power-law distribution. The red progenitor masses and uncertainties are the same in both panels, but the power-law index that the grey lines are drawn from are: \textit{Left:} The best-fitting power-law index of $-2.6^{+0.5}_{-0.6}$; \textit{Right:} The best-fit index from the M83 SNR progenitor sample (-2.9).}
     \label{fig13}
 \end{figure}

Based on the potential bias from gas emission we choose our preferred subsample as the SNRs that did not use Broad-V band data to constrain the progenitor mass as well as the historically observed events.
KS tests on this subsample returned a 90\% confidence range of indices between -3.2 and -2.1 with a peak at -2.6.
Figure \ref{fig13} shows the ranked inferred progenitor masses of this subsample as red marks, with uncertainties, to show the cumulative distribution.
Overplotted are draws from various power-law distributions.
The left panel is the best fitting index while the right panel has draws from the best fit index from \cite{Williams19} (-2.9).
Running KS test between this subsample and the M83 and M31+M33 distributions returned p-values of 0.31 and 0.15, respectively, suggesting there may be a difference in the stellar populations between the galaxies.
Future work will use a more robust approach to determine if this is statistically significant by properly handling the uncertainties in the progenitor mass estimates.

We found that including a contamination CMD in our fitting was critical to obtaining a reliable mass measurement.
In our full catalog, not doing so resulted in significantly fewer Type Ia candidates (16 with and 5 without).
We also found a larger number of progenitor masses between 7.5 and 16 M$_{\odot}$ without fitting for contamination as compared to when a contamination CMD was included.
This suggests that the youngest population present at the SNR location was often under weighted if contamination was not considered.

Our control regions had a significantly different distribution of mass inferences, as well as a higher fraction of region with no young population, than our SNR sample. 
The mass distribution was best matched by a power-law with an index of -3.4 (90\% confidence between -3.7 and -3.1), considerably steeper than the index of our preferred sample.
The fraction of locations that showed no SF within the past 50 Myr was over twice as high as our SNRs.
The amount of smaller inferred progenitor masses indicates that less SF has taken place in these regions within the past 50 Myr as compared to our catalog.
The difference in power-law indices suggests that our SNRs come from a distribution that differs from the general field.

\section{SUMMARY} \label{sect5}

We used $HST$ images to measure the stellar populations surrounding \replaced{189}{185} SNRs, \replaced{10}{9} historically observed SNe, and the black hole formation candidate NGC6946-BH1.
Using CMDs of the stars within 50 pc of each location, we were able to determine the SFH distribution over the past 50 Myr.
From these we inferred the most likely progenitor mass for \replaced{175}{169} remnants.
The remaining 16 showed no sign of excess SF as compared to the general field, making them good Type Ia candidates.

From the 37 remnants with data in three filters, we were able to determine that Broad-V filters appear to bias age estimate for some SNR locations.
At the distance of NGC 6946, emission from the expanding envelope and surrounding interstellar medium can affect the measured photometry.
Stars residing within these pockets of emission appear brighter in Broad-V filters.
As a result, their photometry is biased in a way that causes them to appear extremely red in B$-$Broad-V CMDs, which can cause them to be fit by older populations, and very blue in Broad-V$-$I CMDs, causing MATCH to fit them as younger populations.
By looking at the locations observed in filters that exclude gas emission contamination, we were able to optimize our sample for statistics, resulting in a power-law index of $-2.6^{+0.5}_{-0.6}$.

Our distribution is similar to M31+M33 \citep{Diaz18} and M83 \citep{Williams19} in that few of the progenitor masses are $>$20 M$_{\odot}$.
However, nearly 24\% of the progenitors in our preferred sample have masses $>$20 M$_{\odot}$.
The large fraction can be attributed to the small number of locations in the sample.
\cite{Diaz18} found only around 5\% of progenitors had masses $>$20 M$_{\odot}$, which is reflected in their power-law index of -4.2.
\cite{Williams19} found around 14\% of their progenitors had masses $>$20 M$_{\odot}$, but had significantly more progenitors in their sample leading to a steeper power-law index (-2.9).

To better understand the SNR population within NGC 6946 more data in bands that do not include H$\alpha$, [N II], and [S II] are required.
Photometry with the required precision and depth is currently lacking for NGC 6946 in filters not contaminated by nearby gas for most SNR locations.
Studies in other galaxies, such as M83, have been successful using near-ultraviolet filters in combination with both B and I filters.
Having a $HST$ broad I band survey would be instrumental in constraining the progenitor masses within the galaxy in the future.
Since the NGC 6946 has been the host to more observed SNe than any other, learning more about its SNR population is paramount to better our understanding of SNe in general.

Support for this work was provided by NASA through grants GO-15216, GO-14786, and AR-15042 from the Space Telescope Science Institute, which is operated by AURA, Inc., under NASA contract NAS 5-26555.
WPB acknowledges grant HST-GO-15216-A to Johns Hopkins University, and for partial support from the JHU Center for Astrophysical Sciences.

\textit{Software:} Astrodrizzle \citep{Hack12}, DOLPHOT \citep{Dolphin00,Dolphin16}, PyRAF \citep{SSB12}, TinyTim PSFs \citep{Krist11}, MATCH \citep{Dolphin02,Dolphin12,Dolphin13}.

\bibliographystyle{aasjournal}

\begin{thebibliography}

\bibitem[Adams et al.(2017)]{Adams17}
Adams, S. M., Kochanek, C. S., Gerke, J. R., Stanek, K. Z., \& Dai, X. 2017, MNRAS, 468, 4968, doi: 10.1093/mnras/stx816

\bibitem[Adams et al.(2016)]{Adams16}
Adams, S. M., Kochanek, C. S., Prieto, J. L., et al. 2016, MNRAS, 460, 1645, doi: 10.1093/mnras/stw1059

\bibitem[Anand et al.(2018)]{Anand18}
Anand, G. S., Rizzi, L., \& Tully, R. B. 2018, ApJ, 156, 105, doi: 10.3847/1538-3881/aad3b2

\bibitem[Anderson \& Soto(2013)]{AS13}
Anderson, J. P., \& Soto, M. 2013, A\&A, 550, A69, doi: 10.1051/0004-6361/201220600

\bibitem[Auchettl et al.(2019)]{Auchettl19}
Auchettl, K., Lopez, L. A., Badenes, C., et al. 2019, ApJ,871, 64, doi: 10.3847/1538-4357/aaf395

\bibitem[Badenes(2009)]{Badenes09}
Badenes, C., Harris, J., Zaritsky, D., \& Prieto, J. L. 2009,ApJ, 700, 727, doi: 10.1088/0004-637X/700/1/727

\bibitem[Bastian \& Goodwin(2006)]{BG06}
Bastian, N., \& Goodwin, S. P. 2006, MNRAS, 369, L9, doi: 10.1111/j.1745-3933.2006.00162.x

\bibitem[Blair et al.(2001)]{Blair01}
Blair, W. P., Fesen, R. A., \& Schlegel, E. M. 2001, AJ, 121, 1497, doi: 10.1086/319426

\bibitem[Botticella et al.(2009)]{Botticella09}
Botticella, M. T., Pastorello, A., Smartt, S. J., et al. 2009, MNRAS, 398, 1041, doi: 10.1111/j.1365-2966.2009.15082.x

\bibitem[Bressan et al.(2012)]{Bressan12}
Bressan, A., Marigo, P., Girardi, L., et al. 2012, MNRAS, 427, 127, doi: 10.1111/j.1365-2966.2012.21948.x

\bibitem[Cedr{\'e}s et al.(2012)]{Ced12}
Cedr{\'e}s, B., Cepa, J., Bongiovanni, {\'A}., et al. 2012, A\&A, 545, A43, doi: 10.1051/0004-6361/201219571

\bibitem[Davies \& Beasor(2020)]{DB20}
Davies, B., \& Beasor, E. R. 2020, MNRAS, 493, 468, doi: 10.1093/mnras/staa174

\bibitem[de Blok et al.(2008)]{Blok08}
de Blok, W. J. G., Walter, F., Brinks, E., et al. 2008, AJ, 136, 2648, doi: 10.1088/0004-6256/136/6/2648

\bibitem[D{\'i}az-Rodr{\'i}guez et al.(2018)]{Diaz18}
D{\'i}az-Rodr{\'i}guez, M., Murphy, J. W., Rubin, D. A., et al. 2018, ApJ, 861, 92, doi: 10.3847/1538-4357/aac6e1

\bibitem[D{\'i}az-Rodr{\'i}guez et al.(2021)]{Diaz21}
D{\'i}az-Rodr{\'i}guez, M., Murphy, J. W., Williams, B. F., Dalcanton, J. J., \& Dolphin, A. E. 2021, arXiv e-prints, arXiv:2101.11012

\bibitem[Dolphin(2000)]{Dolphin00}
Dolphin, A. E. 2000, PASP, 112, 1383, doi: 10.1086/316630

\bibitem[Dolphin(2002)]{Dolphin02}
---. 2002, MNRAS, 332, 91, doi: 10.1046/j.1365-8711.2002.05271.x

\bibitem[Dolphin(2012)]{Dolphin12}
---. 2012, ApJ, 751, 60, doi: 10.1088/0004-637X/751/1/60

\bibitem[Dolphin(2013)]{Dolphin13}
---. 2013, ApJ, 775, 76, doi: 10.1088/0004-637X/775/1/76

\bibitem[Dolphin(2016)]{Dolphin16}
---. 2016, DOLPHOT: Stellar photometry. http://ascl.net/1608.013

\bibitem[Dunne et al.(2000)]{Dunne00}
Dunne, B. C., Gruendl, R. A., \& Chu, Y.-H. 2000, AJ, 119,1172, doi: 10.1086/301264

\bibitem[Eldridge \& Xiao(2019)]{Eldridge19}
Eldridge, J. J., \& Xiao, L. 2019, Monthly Notices of the Royal Astronomical Society: Letters, 485, L58–L61, doi: 10.1093/mnrasl/slz030

\bibitem[Evans et al.(2010)]{Evans10}
Evans, I. N., Primini, F. A., Glotfelty, K. J., et al. 2010, ApJS, 189, 37, doi: 10.1088/0067-0049/189/1/37

\bibitem[Fraser et al.(2011)]{Fraser11}
Fraser, M., Ergon, M., Eldridge, J. J., et al. 2011, MNRAS, 417, 1417, doi: 10.1111/j.1365-2966.2011.19370.x

\bibitem[Gal-Yam et al.(2007)]{GalYam07}
Gal-Yam, A., Leonard, D. C., Fox, D. B., et al. 2007, ApJ, 656, 372, doi: 10.1086/510523

\bibitem[Gerke et al.(2015)]{Gerke15}
Gerke, J. R., Kochanek, C. S., \& Stanek, K. Z. 2015, MNRAS, 450, 3289, doi: 10.1093/mnras/stv776

\bibitem[Girardi et al.(2002)]{Girardi02}
Girardi, L., Bertelli, G., Bressan, A., et al. 2002, A\&A, 391, 195, doi: 10.1051/0004-6361:20020612

\bibitem[Gogarten et al.(2009)]{Gogarten09}
Gogarten, S. M., Dalcanton, J. J., Murphy, J. W., et al. 2009, ApJ, 703, 300, doi: 10.1088/0004-637X/703/1/300

\bibitem[Gossage et al.(2018)]{Gossage18}
Gossage, S., Conroy, C., Dotter, A., et al. 2018, ApJ, 863,67, doi: 10.3847/1538-4357/aad0a0

\bibitem[Guillochon et al.(2017)]{Guillochon17}
Guillochon, J., Parrent, J., Kelley, L. Z., \& Margutti, R. 2017, ApJ, 835, 64, doi: 10.3847/1538-4357/835/1/64

\bibitem[Gusev et al.(2013)]{Gusev17}
Gusev, A. S., Sakhibov, F. H., \& Dodonov, S. N. 2013, Astrophysical Bulletin, 68, 40, doi: 10.1134/S1990341313010045

\bibitem[Hack et al.(2012)]{Hack12}
Hack, W. J., Dencheva, N., Fruchter, A. S., et al. 2012, in American Astronomical Society Meeting Abstracts, Vol.220, American Astronomical Society Meeting Abstracts \#220, 135.15

\bibitem[Holoien et al.(2019)]{Holoien19}
Holoien, T. W. S., Brown, J. S., Vallely, P. J., et al. 2019, MNRAS, 484, 1899, doi: 10.1093/mnras/stz073

\bibitem[Jarrett et al.(2013)]{Jarrett13}
Jarrett, T. H., Masci, F., Tsai, C. W., et al. 2013, AJ, 145,6, doi: 10.1088/0004-6256/145/1/6

\bibitem[Jennings et al.(2012)]{Jennings12}
Jennings, Z. G., Williams, B. F., Murphy, J. W., et al. 2012, ApJ, 761, 26, doi: 10.1088/0004-637X/761/1/26

\bibitem[Jennings et al.(2014)]{Jennings14}
---. 2014, ApJ, 795, 170, doi: 10.1088/0004-637X/761/1/26

\bibitem[Kilpatrick \& Foley(2018)]{KF18}
Kilpatrick, C. D., \& Foley, R. J. 2018, MNRAS, 481, 2536, doi: 10.1093/mnras/sty2435

\bibitem[Kilpatrick et al.(2021)]{Kilpatrick21}
Kilpatrick, C. D., Drout, M. R., Auchettl, K., et al. 2021, arXiv e-prints, arXiv:2101.03206.

\bibitem[Krist et al.(2011)]{Krist11}
Krist, J. E., Hook, R. N., \& Stoehr, F. 2011, in Society of Photo-Optical Instrumentation Engineers (SPIE) Conference Series, Vol. 8127, Proc. SPIE, 81270J, doi: 10.1117/12.892762

\bibitem[Lennarz et al.(2012)]{Lennarz12}
Lennarz, D., Altmann, D., \& Wiebusch, C. 2012, A\&A,538, A120, doi: 10.1051/0004-6361/201117666

\bibitem[Li et al.(2006)]{Li06}
Li, W., Van Dyk, S. D., Filippenko, A. V., et al. 2006, ApJ, 641, 1060, doi: 10.1086/499916

\bibitem[Long et al.(2020)]{Long20}
Long, K. S., Blair, W. P., Winkler, P. F., \& Lacey, C. K.2020, arXiv e-prints, arXiv:2007.01415.

\bibitem[Long et al.(2019)]{Long19}
\defcitealias{Long19}{L19}
Long, K. S., Winkler, P. F., \& Blair, W. P. 2019, ApJ, 875, 85, doi: 10.3847/1538-4357/ab0d94

\bibitem[Murphy et al.(2011)]{Murphy11} 
Murphy, J. W., Jennings, Z. G., Williams, B., Dalcanton,J. J., \& Dolphin, A. E. 2011, ApJL, 742, L4, doi:

\bibitem[Murphy et al.(2018)]{Murphy18} 
Murphy, J. W., Khan, R., Williams, B., et al. 2018, ApJ,860, 117, doi: 10.3847/1538-4357/aac2be

\bibitem[O'Neill et al.(2019)]{ONeill19} 
O’Neill, D., Kotak, R., Fraser, M., et al. 2019, A\&A, 622, L1, doi: 10.1051/0004-6361/201834566

\bibitem[Pejcha \& Thompson(2015)]{PT15} 
Pejcha, O., \& Thompson, T. A. 2015, ApJ, 801, 90, doi: 10.1088/0004-637X/801/2/90

\bibitem[Prieto et al.(2008)]{Prieto08} 
Prieto, J. L., Kistler, M. D., Thompson, T. A., et al. 2008, ApJL, 681, L9, doi: 10.1086/589922

\bibitem[Prieto et al.(2010)]{Prieto10} 
Prieto, J. L., Szczygiel, D. M., Kochanek, C. S., et al. 2010, arXiv e-prints, arXiv: 1007.0011.

\bibitem[Salpeter(1955)]{Salpeter55} 
Salpeter, E. E. 1955, ApJ, 121, 161, doi: 10.1086/145971

\bibitem[Schlafly \& Finkbeiner(2011)]{SF11} 
Schlafly, E. F., \& Finkbeiner, D. P. 2011, ApJ, 737, 103, doi: 10.1088/0004-637X/737/2/103

\bibitem[Science Software Branch at STScI.(2012)]{SSB12} 
Science Software Branch at STScI. 2012, PyRAF: Python alternative for IRAF. https://ascl.net/1207.011

\bibitem[Senchyna et al.(2015)]{Senhyna15}
Senchyna, P., Johnson, L. C., Dalcanton, J. J., et al. 2015, ApJ, 813, 31, doi: 10.1088/0004-637X/813/1/31

\bibitem[Skillman et al.(2017)]{Skillman17}
Skillman, E. D., Monelli, M., Weisz, D. R., et al. 2017, ApJ, 837, 102, doi: 10.3847/1538-4357/aa60c5

\bibitem[Smartt(2015)]{Smartt15}
Smartt, S. J. 2015, PASA, 32, e016, doi: 10.1017/pasa.2015.17

\bibitem[Smartt et al.(2009)]{Smartt09}
Smartt, S. J., Eldridge, J. J., Crockett, R. M., \& Maund, J. R. 2009, MNRAS, 395, 1409, doi: 10.1111/j.1365-2966.2009.14506.x

\bibitem[Smartt et al.(2003)]{Smartt03}
Smartt, S. J., Maund, J. R., Gilmore, G. F., et al. 2003, MNRAS, 343, 735, doi: 10.1046/j.1365-8711.2003.06718.x

\bibitem[Smartt et al.(2004)]{Smartt04}
Smartt, S. J., Maund, J. R., Hendry, M. A., et al. 2004, Science, 303, 499, doi: 10.1126/science.1092967

\bibitem[Steele et al.(2008)]{Steele08}
Steele, T. N., Silverman, J. M., Ganeshalingam, M., et al. 2008, Central Bureau Electronic Telegrams, 1275, 1

\bibitem[Sukhbold et al.(2016)]{Sukhbold16}
Sukhbold, T., Ertl, T., Woosley, S. E., Brown, J. M., \& Janka, H. T. 2016, ApJ, 821, 38, doi: 10.3847/0004-637X/821/1/38

\bibitem[Sukhbold \& Woosley(2014)]{SW14}
Sukhbold, T., \& Woosley, S. E. 2014, ApJ, 783, 10, doi: 10.1088/0004-637X/783/1/10

\bibitem[Tominaga et al.(2013)]{Tominaga13}
Tominaga, N., Blinnikov, S. I., \& Nomoto, K. 2013, ApJL, 771, L12, doi: 10.1088/2041-8205/771/1/L12

\bibitem[Van Dyk(2017)]{VD17}
Van Dyk, S. D. 2017, Philosophical Transactions of the Royal Society of London Series A, 375, 20160277, doi: 10.1098/rsta.2016.0277

\bibitem[Van Dyk et al.(2017)]{VDe17}
Van Dyk, S. D., Filippenko, A. V., Fox, O. D., et al. 2017, The Astronomer’s Telegram, 10378, 1

\bibitem[Van Dyk et al.(2003)]{VD03}
Van Dyk, S. D., Li, W., \& Filippenko, A. V. 2003, PASP, 115, 1, doi: 10.1086/345748

\bibitem[Van Dyk et al.(2018)]{VD18}
Van Dyk, S. D., Zheng, W., Brink, T. G., et al. 2018, ApJ, 860, 90, doi: 10.3847/1538-4357/aac32c

\bibitem[Van Dyk et al.(2019)]{VD19}
Van Dyk, S. D., Zheng, W., Maund, J. R., et al. 2019, ApJ, 875, 136, doi: 10.3847/1538-4357/ab1136

\bibitem[Weisz et al.(2014)]{Weisz14}
Weisz, D. R., Dolphin, A. E., Skillman, E. D., et al. 2014, ApJ, 789, 147, doi: 10.1088/0004-637X/789/2/147

\bibitem[Wiggins(2017)]{Wiggins17}
Wiggins, P. 2017, Transient Name Server Discovery Report, 2017-548, 1

\bibitem[Williams et al.(2009)]{Williams09}
Williams, B. F., Dalcanton, J. J., Dolphin, A. E., Holtzman, J., \& Sarajedini, A. 2009, ApJL, 695, L15, doi: 10.1088/0004-637X/695/1/L15

\bibitem[Williams et al.(2017)]{Williams17}
Williams, B. F., Dolphin, A. E., Dalcanton, J. J., et al. 2017, ApJ, 846, 145, doi: 10.3847/1538-4357/aa862a

\bibitem[Williams et al.(2019)]{Williams19}
Williams, B. F., Hillis, T. J., Blair, W. P., et al. 2019, ApJ, 881, 54, doi: 10.3847/1538-4357/ab2190

\bibitem[Williams et al.(2018)]{Williams18}
Williams, B. F., Hillis, T. J., Murphy, J. W., et al. 2018, ApJ, 860, 39, doi: 10.3847/1538-4357/aaba7d

\bibitem[Williams et al.(2014a)]{Williams14a}
Williams, B. F., Peterson, S., Murphy, J., et al. 2014a, ApJ, 791, 105, doi: 10.1088/0004-637X/791/2/105

\bibitem[Williams et al.(2014b)]{Williams14b}
Williams, B. F., Lang, D., Dalcanton, J. J., et al. 2014b, ApJS, 215, 9, doi: 10.1088/0067-0049/215/1/9

\bibitem[Woosley et al.(2002)]{Woosley02}
Woosley, S. E., Heger, A., \& Weaver, T. A. 2002, Reviews of Modern Physics, 74, 1015, doi: 10.1103/RevModPhys.74.1015

\bibitem[Xiao et al.(2019)]{Xiao19}
Xiao, L., Galbany, L., Eldridge, J. J., \& Stanway, E. R. 2019, MNRAS, 482, 384, doi: 10.1093/mnras/sty2557

\bibitem[Yoon et al.(2010)]{Yoon10}
Yoon, S. C., Woosley, S. E., \& Langer, N. 2010, ApJ, 725, 940, doi: 10.1088/0004-637X/725/1/940

\end{thebibliography}

\begin{deluxetable}{ccccccccc} 
\label{tab1}
\tablewidth{0pt} 
\tablecolumns{9} 
\tablecaption{NGC 6946 Observations} 
\tablehead{
\colhead{Image Name} & \colhead{PID} & \colhead{Date} & \colhead{RA (J2000)} & \colhead{Dec (J2000)} & \colhead{V3 Axis (deg)} & \colhead{Exp Time (sec)} & \colhead{Camera} & \colhead{Filter} }
\startdata 
jbky01gjq & 12331 & 2010-08-24 & 308.80594 & 60.19238 & 346.99951 & 521 & ACS & F814W \\
jbky01gmq & 12331 & 2010-08-24 & 308.80604 & 60.19222 & 346.99960 & 603 & ACS & F814W \\
jbky01gqq & 12331 & 2010-08-24 & 308.80613 & 60.19206 & 346.99970 & 603 & ACS & F814W \\
jbky01gtq & 12331 & 2010-08-24 & 308.80575 & 60.19233 & 346.99939 & 703 & ACS & F814W \\
jbky02nbq & 12331 & 2011-08-07 & 308.80610 & 60.19232 & 346.99960 & 521 & ACS & F555W \\
jbky02neq & 12331 & 2011-08-07 & 308.80619 & 60.19216 & 346.99960 & 603 & ACS & F555W \\
jbky02niq & 12331 & 2011-08-07 & 308.80629 & 60.19200 & 346.99970 & 603 & ACS & F555W \\
jbky02nlq & 12331 & 2011-08-07 & 308.80591 & 60.19244 & 346.99939 & 703 & ACS & F555W \\
jbok05oaq & 12450 & 2012-08-30 & 308.74999 & 60.20574 & 328.98950 & 521 & ACS & F814W \\
jbok05oeq & 12450 & 2012-08-30 & 308.75025 & 60.20565 & 328.98969 & 600 & ACS & F814W \\
jbok05p2q & 12450 & 2012-08-30 & 308.75010 & 60.20558 & 328.98959 & 600 & ACS & F814W \\
... & ... & ... & ... & ... & ... & ... & ... \\
\enddata 
\tablenotetext{}{Full tables available in supplemental materials.}
\end{deluxetable} 

\begin{deluxetable}{ccccccccc} 
\label{tab2}
\tablewidth{0pt} 
\tablecolumns{9} 
\tablecaption{SNR photometry depth, extinction, and progenitor mass estimates} 
\tablehead{
\colhead{ID} & \colhead{Bluer Filter} & \colhead{50\% Completeness} & \colhead{Redder Filter} & \colhead{50\% Completeness} & \colhead{Count} & \colhead{$dA_{v}$} & \colhead{$A_{v}$} & \colhead{Mass (M$_{\odot}$)} }
\startdata 
L19-002 & WFC606W & 28.5 & WFC814W & 27.6 & 54 & 0.0 & 1.8 & $46.8^{+5.3}_{-37.2}$ \\
L19-003 & WFC606W & 28.6 & WFC814W & 27.7 & 42 & 0.0 & 0.8 & $13.0^{+5.7}_{-2.1}$ \\
L19-005 & WFC606W & 28.4 & WFC814W & 27.6 & 53 & 0.0 & 1.4 & $12.7^{+3.0}_{-3.7}$ \\
L19-011 & WFC435W & 28.2 & WFC814W & 27.1 & 16 & 0.0 & 1.7 & $7.9^{+5.5}_{-0.2}$ \\
L19-012 & WFC435W & 28.2 & WFC814W & 27.1 & 26 & 0.2 & 1.45 & $8.4^{+7.3}_{-0.3}$ \\
L19-013 & WFC606W & 28.6 & WFC814W & 27.7 & 63 & 0.0 & 1.3 & $11.3^{+5.8}_{-1.7}$ \\
L19-014 & WFC435W & 28.2 & WFC814W & 27.1 & 23 & 0.0 & 1.5 & $13.0^{+4.1}_{-4.0}$ \\
L19-016 & WFC435W & 28.1 & WFC606W & 28.0 & 17 & 0.0 & 1.55 & $9.1^{+3.4}_{-1.0}$ \\
L19-017 & WFC435W & 28.1 & WFC606W & 27.6 & 19 & 0.2 & 1.65 & $31.5^{+3.2}_{-22.5}$ \\
L19-018 & WFC435W & 28.2 & WFC814W & 27.1 & 19 & 0.0 & 0.95 & $31.9^{+2.8}_{-22.3}$ \\
... & ... & ... & ... & ... & ... & ... & ... & ... \\
\enddata 
\end{deluxetable} 

\begin{deluxetable}{ccccc} 
\label{tab3}
\centering
\tablewidth{0pt} 
\tablecolumns{5} 
\tablecaption{NGC 6946 Supernova Remnants} 
\setlength{\tabcolsep}{5pt} 
\tablehead{
\colhead{ID$^{a,b}$} & \colhead{Other ID$^{b}$} & \colhead{RA (J2000)$^{a,b}$} & \colhead{Dec (J2000)$^{a,b}$} & \colhead{Use Flag}}
\startdata 
L19-001 & --- & 20:34:15.00 & +60:10:44.3 & o \\ 
L19-002 & --- & 20:34:15.48 & +60:07:31.6 & p,h \\ 
L19-003 & --- & 20:34:15.78 & +60:08:26.0 & p,h \\ 
L19-004 & --- & 20:34:16.41 & +60:08:27.3 & i \\ 
L19-005 & --- & 20:34:16.68 & +60:07:30.8 & p,h \\ 
L19-006 & --- & 20:34:17.54 & +60:10:58.3 & o \\ 
L19-007 & --- & 20:34:17.95 & +60:10:00.4 & o \\ 
L19-008 & --- & 20:34:18.39 & +60:10:47.3 & o \\ 
L19-009 & --- & 20:34:18.84 & +60:11:08.9 & o \\ 
L19-010 & --- & 20:34:19.17 & +60:08:57.5 & n \\ 
L19-011 & --- & 20:34:20.60 & +60:09:06.8 & p \\ 
... & ... & ... & ... & ... \\ 
\enddata 
\tablenotetext{a}{\cite{Long19}}
\tablenotetext{b}{\cite{Long20}}
\end{deluxetable} 

\begin{deluxetable}{cccccccccccccc} 
\label{tab4}
\centering
\tablewidth{0pt} 
\tablecolumns{14} 
\tablecaption{Age Probability Distribution Results for L19-049\tablenotemark{a}} 
\tablehead{
\colhead{T1} & \colhead{T2} & \colhead{SFR (Best)} & \colhead{-err} & \colhead{+err} & \colhead{PDF(Best)} & \colhead{-err} & \colhead{+err} & \colhead{CDF(Best)} & \colhead{CDF(Low)} & \colhead{CDF(High)} & \colhead{M1} & \colhead{M2} }
\startdata 
  4.0 &   4.5 &  0.0000e+00 &  0.0000e+00 &  1.5208e-03 &     0.000 &  0.000 &  0.156 &     0.000 &    0.000 &     0.015 &  52.1 &  66.6 \\
  4.5 &   5.0 &  0.0000e+00 &  0.0000e+00 &  1.4471e-03 &     0.000 &  0.000 &  0.165 &     0.000 &    0.000 &     0.038 &  42.0 &  52.1 \\
  5.0 &   5.6 &  0.0000e+00 &  0.0000e+00 &  1.3836e-03 &     0.000 &  0.000 &  0.175 &     0.000 &    0.000 &     0.060 &  34.7 &  42.0 \\
  5.6 &   6.3 &  0.0000e+00 &  0.0000e+00 &  1.2211e-03 &     0.000 &  0.000 &  0.173 &     0.000 &    0.000 &     0.081 &  29.2 &  34.7 \\
  6.3 &   7.1 &  0.0000e+00 &  0.0000e+00 &  1.2449e-03 &     0.000 &  0.000 &  0.194 &     0.000 &    0.000 &     0.103 &  26.0 &  29.2 \\
  7.1 &   7.9 &  0.0000e+00 &  0.0000e+00 &  1.1461e-03 &     0.000 &  0.000 &  0.199 &     0.000 &    0.000 &     0.123 &  23.1 &  26.0 \\
  7.9 &   8.9 &  0.0000e+00 &  0.0000e+00 &  1.1244e-03 &     0.000 &  0.000 &  0.214 &     0.000 &    0.000 &     0.147 &  20.6 &  23.1 \\
  8.9 &  10.0 &  0.0000e+00 &  0.0000e+00 &  1.3134e-03 &     0.000 &  0.000 &  0.263 &     0.000 &    0.000 &     0.179 &  18.7 &  20.6 \\
 10.0 &  11.2 &  8.0267e-04 &  8.0267e-04 &  5.9284e-04 &     0.083 &  0.083 &  0.274 &     0.083 &    0.063 &     0.260 &  17.1 &  18.7 \\
 11.2 &  12.6 &  0.0000e+00 &  0.0000e+00 &  1.1678e-03 &     0.000 &  0.000 &  0.286 &     0.083 &    0.077 &     0.296 &  15.7 &  17.1 \\
 12.6 &  14.1 &  0.0000e+00 &  0.0000e+00 &  1.1340e-03 &     0.000 &  0.000 &  0.304 &     0.083 &    0.083 &     0.335 &  14.5 &  15.7 \\
 14.1 &  15.8 &  0.0000e+00 &  0.0000e+00 &  1.1831e-03 &     0.000 &  0.000 &  0.338 &     0.083 &    0.083 &     0.379 &  13.4 &  14.5 \\
 15.8 &  17.8 &  0.0000e+00 &  0.0000e+00 &  1.2430e-03 &     0.000 &  0.000 &  0.376 &     0.083 &    0.083 &     0.426 &  12.5 &  13.4 \\
 17.8 &  20.0 &  0.0000e+00 &  0.0000e+00 &  1.3650e-03 &     0.000 &  0.000 &  0.426 &     0.083 &    0.083 &     0.484 &  11.7 &  12.5 \\
 20.0 &  22.4 &  1.8534e-03 &  1.8534e-03 &  0.0000e+00 &     0.380 &  0.380 &  0.453 &     0.463 &    0.268 &     0.651 &  10.9 &  11.7 \\
 22.4 &  25.1 &  0.0000e+00 &  0.0000e+00 &  1.3317e-03 &     0.000 &  0.000 &  0.477 &     0.463 &    0.307 &     0.705 &  10.3 &  10.9 \\
 25.1 &  28.2 &  0.0000e+00 &  0.0000e+00 &  1.5272e-03 &     0.000 &  0.000 &  0.540 &     0.463 &    0.355 &     0.768 &   9.6 &  10.3 \\
 28.2 &  31.6 &  1.8519e-03 &  1.8519e-03 &  0.0000e+00 &     0.537 &  0.537 &  0.342 &     1.000 &    0.684 &     1.000 &   9.0 &   9.6 \\
 31.6 &  35.5 &  0.0000e+00 &  0.0000e+00 &  1.6413e-03 &     0.000 &  0.000 &  0.613 &     1.000 &    0.763 &     1.000 &   8.6 &   9.0 \\
 35.5 &  39.8 &  0.0000e+00 &  0.0000e+00 &  1.3850e-03 &     0.000 &  0.000 &  0.600 &     1.000 &    0.849 &     1.000 &   8.1 &   8.6 \\
 39.8 &  44.7 &  0.0000e+00 &  0.0000e+00 &  1.1501e-03 &     0.000 &  0.000 &  0.583 &     1.000 &    0.967 &     1.000 &   7.7 &   8.1 \\
 44.7 &  50.1 &  0.0000e+00 &  0.0000e+00 &  1.0516e-03 &     0.000 &  0.000 &  0.589 &     1.000 &    1.000 &     1.000 &   7.3 &   7.7 \\
\enddata 
\tablenotetext{a}{Similar tables are available for all fitted SNRs with inferred progenitor masses in the supplemental materials. For a description of each column, see Section \ref{sect3}.}
\end{deluxetable} 

\end{document}